\documentclass[%
 reprint,
floatfix,
superscriptaddress,
showpacs,preprintnumbers,
 amsmath,amssymb,
 aps,
pra,
]{revtex4-2}
\DeclareUnicodeCharacter{2212}{-}
\usepackage{xcolor}
\usepackage{times}
\usepackage{amssymb}
\usepackage{dsfont}
\usepackage{graphicx}
\usepackage{dcolumn}
\usepackage{bm}
\usepackage{epstopdf}
\usepackage{siunitx}
\usepackage[colorlinks]{hyperref}
\hypersetup{colorlinks = true,
linkcolor = blue,
filecolor = blue,
urlcolor = blue,
citecolor = blue}

\begin{document}

\title{Correlated Mott semi-metal in the topological heavy fermion model}

\author{Emile Pangburn}
\email{emile.pangburn@ipht.fr}
\affiliation{Universit\'{e} Paris-Saclay, Institut de Physique Th\'eorique,  CEA, CNRS, F-91191 Gif-sur-Yvette, France}

\author{Igor de Melo Froldi}
\affiliation{Universit\'{e} Paris-Saclay, Institut de Physique Th\'eorique,  CEA, CNRS, F-91191 Gif-sur-Yvette, France}
\affiliation{Instituto de F{\'i}sica, Universidade Federal de Goi{\'a}s, 74.690-900, Goi{\^a}nia-GO, Brazil}

\author{Anurag Banerjee}
\email{anurag.banerjee@ipht.fr}
\affiliation{Universit\'{e} Paris-Saclay, Institut de Physique Th\'eorique,  CEA, CNRS, F-91191 Gif-sur-Yvette, France}

\begin{abstract}
The topological heavy-fermion model provides a minimal framework for describing the coexistence of localized moments and itinerant Dirac electrons in magic-angle twisted bilayer graphene (MATBG). Several analytical and numerical methods have been applied to this model; however, whether they provide a realistic description of MATBG remains incompletely understood. In this work, we develop an Hubbard operator approach that incorporates non-local correlations beyond the single-site limit. We benchmark the approximate calculations against numerically exact determinant quantum Monte Carlo simulations of a lattice-regularized model. We show that commonly used local approximations, such as Hubbard-I, fail to capture the coupling between localized and itinerant degrees of freedom, leading to incorrect spectral properties in the local-moment regime. In contrast, the Hubbard operator method provides a better description of both correlation functions and spectral features over a regime of parameters, in good agreement with exact numerical methods.
\end{abstract}

\maketitle

\section{Introduction}
At half-filling, the charge localization of electrons induced by strong electronic repulsion  is a defining hallmark of the Mott physics and is well described by the lattice Hubbard model~\cite{hubbard1963electron, hubbard1964electron, georges1996dynamical, imada1998metal,arovas2022hubbard,qin2022hubbard}. Recently, this phenomenon has attracted renewed interest for flat bands in twisted materials like  Magic-angle twisted bilayer graphene (MATBG).  It shows semi-metallic conductance near charge neutral point and localized moments in local measurements~\cite{zondiner2020cascade,rozen2021entropic,saito2021isospin}. Explaining the coexistence of these contradictory phenomena poses a significant theoretical and experimental challenge~\cite{datta2023heavy,merino2025interplay,xiao2025interacting,zhang2025heavy}. In particular, the topological nature of the bands obstructs the construction of a conventional lattice model, in contrast to the conventional Hubbard setting with sharply localized orbitals~\cite{kang2018symmetry,song2019all, po2019faithful, ahn2019failure,khalaf2019magic, tarnopolsky2019origin,xie2020nature, song2019all,wu2021chern,bernevig2021twisted, zhou2026symmetric}. Understanding how charge localization emerges in such topological flat-band systems is essential not only for characterizing the correlated insulating states at half-filling~\cite{cao2018correlated,choi2019stm,kerelsky2019stm,xie2019stm,bultinck2020ground,wei2025dirac}, but also for elucidating  the emergent ordered phases that arise upon doping~\cite{cao2018sc,lu2019superconductors,khalaf2021charged,yankowitz2019tuning,stepanov2020untying,saito2020independent,wong2020cascade,zondiner2020cascade,das2021qoscs,sharpe2019ferro,kwan2021kekule,shavit2021theory, jaoui2022strangemetal,wagner2022global,tian2023evidence,kwan2024electron,zhao2025resonating,Doped_SymmetricKondo,wei2025theory,waschitz2026momentum,wei2026tunneling}, many of which may trace their origin back to the underlying Mott physics.   

Two distinct theoretical approaches are common to understand the contrasting features of MATBG phenomenology. The first starts from the Bistritzer-MacDonald (BM) model~\cite{bistritzer2011moire} and focus on the low-energy bands~\cite{Po2018,zhao2025topological, zhao2025resonating, ledwith2025nonlocal,nosov2026controlled}, introducing the interactions completely in the momentum space without introducing any extra degree of freedom. The second approach consists in Wannierizing the model, allowing it to be described  by strongly correlated electrons forming local moments along with itinerant Dirac electrons. The topological heavy-fermion model (THFM) has been proposed as an effective description of correlated phases in topological flat-band materials like MATBG~\cite{song2022magic}.

The advantages of THFM approach is that it allows the use of established embedding techniques, where a single site, describing the correlated electrons is immersed in an effective bath. In this context, slave-rotor theory~\cite{lau2025topological}, Hubbard-I based approximations~\cite{hu2025projected,hu2026twistedbilayergraphenelifetimes,wei2026lifetimespectralfunctiontopological} (in which the self-energy takes its atomic form with perturbative corrections to include non-local contributions), Dynamical Mean Field Theory (DMFT)~\cite{cualuguaru2025obtaining,crippa2025dynamical} where the self-energy includes a dynamical back-reaction from the bath and Cluster Perturbation Theory (CPT)~\cite{vituri2026controlledloopexpansiontopological} have been use to compute the electronic spectral function. 

One regime of particular interest in MATBG is characterized by disordered local moments that hybridize weakly with the conduction electrons. In this regime, local approximations that neglect spatial correlations can produce inaccurate spectral features. To address this issue, we develop the Hubbard-operator (HO) method~\cite{ovchinnikov2004hubbard,mancini2004hubbard}, originally introduced to study the Mott transition in the single-band Hubbard model~\cite{hubbard1963electron,hubbard1964electron,roth1969electron}. We benchmark the resulting spectral functions and correlation functions against numerically exact determinant quantum Monte Carlo (DQMC) simulations of lattice realizations of the same model.

\section{Motivation\label{Sec:Motivations}}
\subsection{Local moment regime}
The effective low-energy physics of MATBG consists of low-energy flat bands and higher-energy dispersive bands~\cite{Mixed_Valence}. In the particle--hole symmetric regime, the low-energy theory is primarily controlled by three parameters: the gap between the (quasi-)flat and dispersive bands, $\Delta$, the flat-band bandwidth, $h_b$, and the Coulomb interaction, $U$. In addition, the dimensionless parameter $s$ characterizes the Berry curvature density.

At half-filling and in the regime $\Delta \gg U \gg h_b$, three temperature regimes emerge. At low temperatures, the system occupies an almost flat band with nontrivial quantum geometry, favoring ferromagnetic order by suppressing double occupancy~\cite{tasaki1998nagaoka}. In MATBG, however, the Berry curvature is concentrated near the $\Gamma$ point~\cite{bistritzer2011moire}. Away from $\Gamma$, the electronic states become increasingly localized and the local moments weakly coupled. As a result, ferromagnetic order develops only below the scale $T \lesssim Us^2$. In the opposite limit, $T \gg U$, charge fluctuations dominate and the notion of local moments breaks down. The regime of interest for MATBG is therefore $U \gg T \gg Us^2$, where local moments form but remain disordered and interact only weakly with the conduction electrons.

The parameter $s$ is controlled by the Fermi velocity $v$ of the dispersive bands. As $s \rightarrow 0$, the system enters a weak-coupling regime in which a perturbative expansion in $s$ becomes controlled~\cite{vituri2026controlledloopexpansiontopological}. If, in addition, $U \ll \gamma$, the correlated moments are expected to decouple and behave as isolated local moments.

However, realistic parameter regimes are far from such idealized limit in which the dispersive bands can be completely eliminated. We consider a model that remains amenable to exact calculations to investigate the properties of local moments across different regimes. In particular, we find that the formation of fully decoupled local moments is more restrictive than suggested by simplified low-energy descriptions. The parameter regime required for such decoupling is different from the ab initio predictions~\cite{song2022magic,cualuguaru2025obtaining}. Our results suggest that realistic MATBG reside in a regime where moment formation and electronic hybridization remain deeply intertwined, motivating a more careful treatment of correlations beyond idealized projected-band limit.

\subsection{The topological heavy fermion model\label{Sec:THFM_Framework}}
The topological heavy fermion model provides a simplified description of the local moment regime. This model consists of local orbitals ($f$) with vanishing kinetic energy and delocalized topological conduction electrons ($c_1$ and $c_2$), with approximately unbounded kinetic energy spectrum. The local moments exhibit an $SU(8)$ symmetry that incorporates spin, valley, and orbital degrees of freedom. The self-energy in the projected flat-band basis acquires a strong momentum dependence which reproduces results obtained within the BM model~\cite{ledwith2025exotic,ledwith2025nonlocal,hu2025projected}. This occurs even though the interactions of the $f$ electrons are purely local and they do not hybridize directly between sites, but only indirectly via the conduction electrons. 

The spectral properties of the THFM have been studied using a range of analytical and numerical methods, including DMFT~\cite{rai2024dynamical,cualuguaru2025obtaining}, the Hubbard-I approximation~\cite{hu2025projected} and its extensions~\cite{hu2026twistedbilayergraphenelifetimes,wei2026tunneling}, and cluster perturbation theory~\cite{vituri2026controlledloopexpansiontopological}. Previous work has identified significant limitations of Hubbard-I-based approaches away from the projected limit~\cite{harris1967single,roth1969electron,grober2000anomalous,mancini2004hubbard}. As shown in App.~\ref{Sec:Benchmark_Dimer} for the Hubbard dimer, the Hubbard-I approximation effectively corresponds to an infinite-temperature treatment of correlation functions, where different lattice sites are completely decoupled. CPT~\cite{senechal2000spectral,senechal2002cluster} relies on a similar decoupling and neglects inter-cluster correlations beyond the one-body level. A more systematic treatment is provided by the composite operator method (COM)~\cite{mancini2004hubbard}, which incorporates these inter-cluster correlations.

Similar limitations arise in single-site DMFT, which becomes exact only in the infinite-dimensional limit~\cite{georges1996dynamical} and neglects spatially nonlocal correlations. Consequently, the spectrum is largely determined by the local density of states of the effective bath and fail to capture effects associated with nontrivial quantum geometry. Significant deviations from DMFT spectrum have been observed in low-dimensional systems~\cite{kohno2022emergence,assaad1999quantum,capponi2001spin}.

Recent studies have shown that the electronic self-energy away from the Dirac point grows approximately linearly with the number of flavors~\cite{vituri2026controlledloopexpansiontopological,hu2026twistedbilayergraphenelifetimes,wei2026tunneling}. Further study also suggests that nonlocal correlations play an important role in MATBG~\cite{nosov2026controlled}. Motivated by this observation, we develop an extension of the Hubbard-I approximation that incorporates nonlocal correlations, which we refer to as the Hubbard operator method. A related framework was recently proposed in Ref.~\cite{ma2026green}.

The lattice formulation of the model allows us to benchmark the HO method against numerically exact DQMC simulations. We focus on the less correlated $N_f=2$ case, where the Hubbard bands are expected to be the sharpest. Nevertheless, we find that incorporating nonlocal correlations substantially reduces the gap compared to predictions from local approximations.
\section{Hubbard operator method\label{Sec:OperatorExpansion}}
\subsection{Operator expansion}
The Hubbard-operator method is an analytical quantum-cluster approach~\cite{maier2005quantum} to evaluate the correlation functions. Here we focus on the two-point Greens function (GF) of a Hubbard type Hamiltonian. The method has been reviewed recently in Refs. \cite{haurie2024bands,Pangburn_Layer,banerjee2025probing,Pangburn_Ringstates} and here we outline it briefly for the context. 

A local part of the Hamiltonian  $H_{\mathrm{loc},i}$  is treated exactly leading to the definition of a local basis. The non-local processes couples the different local basis. 
\begin{align}
H = \sum_i H_{\mathrm{loc},i} + \sum_{ij} H_{t,ij},
\end{align}
where $H_{\mathrm{loc},i}$ acts on a finite-dimensional local Hilbert space $\mathcal{H}_i$, and $H_{t,ij}$ couples different clusters $i$ and $j$.

Because $\mathcal{H}_i$ is finite-dimensional, a finite operator basis $\mathbf{\Psi}_{i}$ closes under the equation of motion with the local Hamiltonian
\begin{align}
\mathbf{J}_{\mathrm{loc},i} = [\mathbf{\Psi}_{i},H_{\mathrm{loc},i}]
= E_{\mathrm{loc},i}\mathbf{\Psi}_{i},
\end{align}
where $E_{\mathrm{loc},i}$ is a matrix whose eigenvalues describe the local excitation energies. The intercluster terms generates additional operators by  the equations of motion of local operators,
\begin{align}
[\boldsymbol{\Psi}_i,H_t] = \delta \mathbf{J}_i ,
\end{align}
where $\delta\mathbf{J}_i$ contains operators outside the chosen basis.
The Hubbard-operator method assumes that the chosen basis already captures the relevant quasiparticles and the underlying physics of the model. One therefore projects $\delta\mathbf{J}_i$ back onto the local basis
\begin{align}
\delta \mathbf{J}_i \approx \sum\limits_j E_{t,ij}\boldsymbol{\Psi}_{j} .
\end{align}
The local Hamiltonian therefore determines the quasiparticle content, while the nonlocal terms govern their propagation and hybridization. Combining the local and non-local contributions yields
\begin{align}
&\mathbf{J}_i \approx \mathbf{J}_{\mathrm{loc},i} + \delta\mathbf{J}_{i}
= \sum\limits_{j}( \delta_{ij}E_{\mathrm{loc},i} + E_{t,ij})\boldsymbol{\Psi}_{j}.&
\label{eq:E_matrix_def}
\end{align}

Evaluating $E_{\mathrm{loc}}+E_t$ requires two matrices defined by expectation values:
\begin{align}
&M_{ij} = \langle\{ {\mathbf{J}_i, \boldsymbol{\Psi}^\dagger_{j}} \}\rangle &\\
&I_{ij}= \langle \{ \boldsymbol{\Psi}_{i}, \boldsymbol{\Psi}^\dagger_{j}\}\rangle&
\end{align}
where all commutators are evaluated at equal times. The quasiparticle energy matrix is then given by~\cite{haurie2024bands,mancini2004hubbard}
\begin{align}
E = MI^{-1}.
\end{align}

Both $M$ and $I$ depend on expectation values that must be determined self-consistently. The $E$-matrix encodes the excitation energies, while the $I$-matrix determines their spectral weights because of the non-canonical operator algebra.

The retarded Green’s function for the operator basis is
\begin{align}
\mathds{G}_{ij}(t)
= -i\theta(t)\langle\{\boldsymbol{\Psi}_{i}(t),
\boldsymbol{\Psi}^\dagger_{j}(0)\}\rangle .
\end{align}
Using Eq.~\eqref{eq:E_matrix_def}, one obtains in frequency space:
\begin{align}
&\mathds{G}_{ij}(\omega)
= \left[(\omega+i\varepsilon - E)^{-1}I\right]_{ij},&
\label{eq:GF_def}
\end{align}
where $\varepsilon$ is a phenomenological quasiparticle broadening. Note that the non-canonical matrix $I$  distinguishes the interacting model from a free-fermion case.

\subsection{Cluster Hubbard operator}
In this section, we discuss the choice of cluster, and therefore the operator basis, used within the Hubbard-operator method. The selected basis determines which correlations and excitations are treated exactly and consequently constrains the physics that can be captured within the approximation. In the limit of an infinitely large cluster, the method becomes exact and recovers the full excitations.

Different cluster choices naturally emphasize different physical regimes. A single-site basis favors a local-moment description, as it treats on-site correlations exactly while nonlocal processes are incorporated only through the projection procedure. In contrast, a basis constructed from dimer (two coupled sites) explicitly incorporates short-range hybridization and therefore provides a more natural description of Kondo-like physics. The underlying assumption of quantum-cluster approaches is that the essential low-energy excitations of the model are predominantly local or short-ranged. Consequently, a relatively small cluster is often sufficient to capture the relevant physics, while larger clusters systematically improve the description by incorporating longer-range correlations.

At any finite truncation order, the HO method describes quasiparticles with infinite lifetime. In principle, one can introduce a self-energy for the Hubbard operators themselves, thereby generating finite lifetimes~\cite{mancini2004hubbard}. However, except in special cases, such a procedure is not controlled~\cite{wei2026lifetimespectralfunctiontopological}. Instead, as the operator basis is enlarged, the electron hybridizes with an increasing number of composite excitations and generates an effective electronic self-energy and, consequently, a finite quasiparticle lifetime. 

\subsubsection{Single-site cluster -- Composite operator method}
\paragraph{Operators} The single-dot cluster contains the operators to fully describe the charge excitations of the single-site Hubbard model. The local Hamiltonian is just the interaction part
\begin{align}
&H_{\text{dot}}=U\left(\hat{n}_{1\uparrow}-1/2\right)\left(\hat{n}_{1\downarrow}-1/2\right)&
\end{align}
The dot basis is composed of the holons $\xi_{1\sigma}$ and doublons $\eta_{1\sigma}$, operators given by (in the paramagnetic limit)
\begin{align}\label{BasisParamag}
&\mathbf{\Psi} = \left(\xi_{1\uparrow},\eta_{1\uparrow}\right)^T&
\end{align}
defined as
\begin{align}
\xi_{1\sigma} = c_{1\sigma}(1-n_{1\bar\sigma}),\qquad
\eta_{1\sigma} = c_{1\sigma}n_{1\bar\sigma},
\end{align}
The electron operator can be decomposed as a function of the holon and doublon operators
\begin{align}
c_{1\sigma} = \xi_{1\sigma} + \eta_{1\sigma}.
\end{align}
The electronic retarded GF $\mathcal{G}$ then follows from the components of $\mathds{G}$:
\begin{align}
&\mathcal{G}(\omega)= \mathds{G}_{\xi\xi}(\omega)+\mathds{G}_{\xi\eta}(\omega)+\mathds{G}_{\eta\xi}(\omega)+\mathds{G}_{\eta\eta}(\omega)&
\end{align}

The single-site basis for the model studied in this work is presented in App.~\ref{App:ReviewCOM}.

\paragraph{Correlation function expectation} 
The equation of motion for the holon-doublon basis closes for the local Hubbard type interaction in the atomic limit. The hopping terms generates additional operators that are outside the basis. The projection to the chosen operator basis connects the operators at different sites $\alpha$ and $\alpha'$, with $\alpha \neq \alpha'$, generating non-local correlations. This coupling is mainly characterized by a parameter $p$, expressed as:
\begin{align}
&p_{\alpha \alpha'}=\langle n_{\alpha\downarrow}n_{\alpha' \downarrow}\rangle +\langle S_{\alpha}^-S_{\alpha'}^+ \rangle -\langle \Delta_{\alpha} \Delta_{\alpha'}^\dagger\rangle,   \label{eq:pij}&
\end{align}
which depends on density-density, spin-spin and pair-pair correlations.

In the composite operator method~\cite{mancini2004hubbard}, the $p$-parameter is evaluated self-consistently using Roth decoupling scheme to express it in terms of two-point correlations (for details See Refs.~\cite{haurie2024bands,banerjee2025charge}). Therefore for COM, the correlations are functions of temperature $T$ and interaction strength.

In contrast, the Hubbard I approximation corresponds to a high-temperature limit of the COM in which inter-site correlations are neglected. Therefore, the $p$ term becomes
$\langle n_{\alpha \sigma}n_{\alpha' \sigma'}\rangle = \langle n_{\alpha \sigma} \rangle \langle n_{\alpha' \sigma'} \rangle$, $\langle S_\alpha^-S_{\alpha'}^+ \rangle = 0$ and $\langle\Delta_{\alpha}\Delta_{\alpha'}^\dagger\rangle = 0$. Within this approximation, the parameter $p$ is interaction strength and temperature independent, and therefore fixed at half-filling to $p = 0.25$. As the temperature increases above the interaction strength $U$, the COM correlations asymptotically approaches the Hubbard I correlations as shown in App.~\ref{App:Corr_Temperature}. 

\subsubsection{Dimer basis}
The solution of the single Hubbard dimer using the equation of motion technique is outlined in Refs.~\cite{silant2015dimer,catalano2018hubbard,mironov2025dimer}. However, the cluster Hubbard operator calculations for a dimer basis has never performed to the best of our knowledge~\cite{hubbard1963electron,hubbard1964electron}.

\paragraph{Operators} The dimer basis consists of the operators required to fully describe the charge excitations of the two-site Hubbard model defining the local Hamiltonian,
\begin{align}
H_{\text{d}}=&-t\sum\limits_\sigma\left(c^\dagger_{1\sigma}c_{2\sigma}+c^\dagger_{2\sigma}c_{1\sigma}\right)\\
\nonumber&+\sum\limits_{l=1,2}U_l\left(\hat{n}_{l\downarrow}-1/2\right)\left(\hat{n}_{l\uparrow}-1/2\right)
\end{align}

Te following basis closes under the equation of motion of the dimer Hamiltonian
\begin{align}
\nonumber&\mathbf{\Psi}_{\text{d}}=\big(\xi_{1\uparrow},\eta_{1\uparrow},\xi_{1\uparrow}n_{2\downarrow},\eta_{1\uparrow}n_{2\downarrow},\xi_{1\uparrow}n_{2\uparrow},\eta_{1\uparrow}n_{2\uparrow},\xi_{1\downarrow}S^-_{2}, &\\
&\eta_{1\downarrow}S^-_{2},\xi^\dagger_{1\downarrow}\Delta_{2},\eta^\dagger_{1\downarrow}\Delta_{2},\xi_{1\uparrow}n_{2\uparrow}n_{2\downarrow},\eta_{1\uparrow}n_{2\uparrow}n_{2\downarrow},(1\leftrightarrow 2)\big)^T&
\end{align}

The electronic GF can be obtained in the same way as the single-dot basis. 

\paragraph{Correlation function expectation}
The $M$ and $I$ matrices in the dimer basis are functions of expectation values and must be evaluated to extract spectral properties. However, due to the large number of operators in the dimer basis, implementing a Roth-type decoupling~\cite{roth1969electron} is tedious. Instead, we obtain the required correlation functions using finite-temperature DQMC simulations as implemented in the \textit{Algorithms for Lattice Fermions} (ALF) library~\cite{bercx2017alf,ALFcode}.

\section{Model\label{Sec:Model}}
As discussed in Sec.~\ref{Sec:THFM_Framework}, the model describing MATBG is the THFM with $SU(8)$ symmetry. The presence of additional internal quantum numbers, combined with the long-range hopping of the delocalized electrons (single-Dirac cone) restricts the analysis of the THFM. Assessing the validity of the physical picture emerging from such approximate treatments, and comparing it with numerically exact methods, requires a lattice formulation that captures the essential physical features of the continuum model near certain high-symmetry momentum points. 

We study a lattice regularized version of the topological heavy-fermion model, given by:

\begin{align}
&\mathcal{H}_{\rm THFM} = \frac{U}{2} \sum\limits_i\left[\sum_{\sigma=1}^{N_f} \left( f^{\dagger}_{i\sigma} f_{i\sigma} -\frac{1}{2}\right)\right]^2& \nonumber \\
&-\gamma \sum\limits_{i\sigma}\left[f_{i\sigma}^\dagger c_{1i\sigma}+h.c.\right] -\sum\limits_{\langle ij\rangle \sigma}\left[t_{ij}c^\dagger_{1i\sigma}c_{2j\sigma} + h.c.\right],
\label{eq:THFM1_model}
\end{align}
where $U$ is the on-site repulsion on the $f$-orbitals. The hopping amplitude is $t_{jl}=t(\delta_{l,j+1}  - \delta_{l,j-1})$, generating a non-interacting Dirac like dispersion $\epsilon(k)\sim\sin(k)$. This dispersion hosts a linear, gapless crossing at $k=0,\pm \pi$, reproducing the low-energy structure of the continuum topological heavy-fermion model near the $\Gamma$ point. The parameter $\gamma$ is the hybridization between $f$ and $c_1$ electrons.

We study the model in one dimension to provide an exact benchmark against DQMC in the thermodynamic limit. Furthermore, in $1D$ long-range magnetic order is absent so the paramagnetic HO approximation, Eq. \eqref{BasisParamag}, is expected to remain accurate down to lowest temperatures. 

\subsection{Band structure continuum model\label{Sec:LatticeContinuum}}
\begin{figure}[h!]
    \centering
    \includegraphics[width=1.\linewidth]{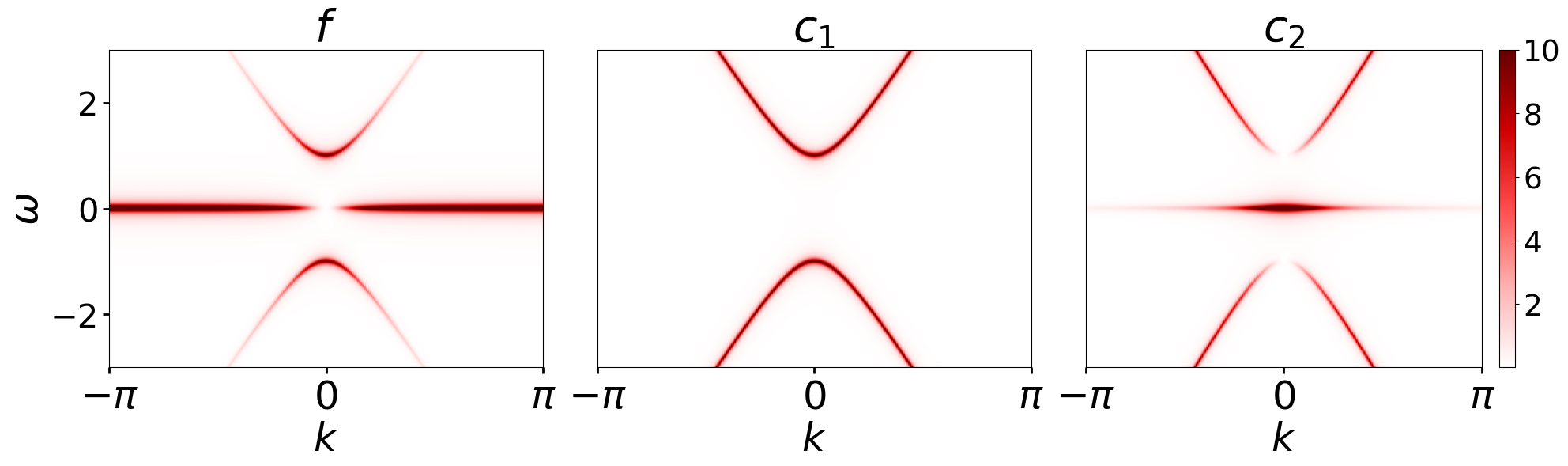}
    \caption{Spectral function of the correlated $f$ electrons (left) and the two itinerants electrons $c_1$ and 
    $c_2$ (middle and right) in the non-interacting continuum topological heavy fermion model (with $\sin(k)\rightarrow k$). 
    The parameters used are $\gamma = 1.0$, $t=1.0$ and quasiparticle broadening $\varepsilon=0.05$.  }
    \label{fig:fig1}
\end{figure}

Fig.~\ref{fig:fig1} shows the orbital-resolved spectrum of the continuum topological heavy-fermion model. The key features are as follows: the $c_1$ electrons are fully gapped, while a perfectly flat band remains pinned at $\omega=0$. Near the $\Gamma$ point, this flat band is dominated by $c_2$ character, whereas away from $\Gamma$ it becomes predominantly $f$-like. This structure originates from a Dirac cone formed by the $c_1$–$c_2$ sector, which is partially gapped by hybridization with the $f$ electrons. Upon introducing interactions, the flat band is expected to remain pinned at $\omega=0$ close to $\Gamma$, where it is dominated by the non-interacting $c_2$ orbital, while at larger momenta an interaction-induced splitting of order $U$ should develop because of the dominance of $f$-electrons. In the following, we denote $W_2$ the bandwidth of the non-interacting $c_2$ electrons, $W_{2}\equiv 4t$. This is the equivalent of $v\Lambda_c$ of the continuum model~\cite{song2022magic}.

\subsection{Basis used}
We use the single-dot and dimer basis for the THFM model.

\paragraph{Single dot basis}

The single-dot basis used is the following 
\begin{align}
&\mathbf{\Psi}_{i\sigma} = \left(\xi_{fi\sigma},\eta_{fi\sigma},\xi_{c_1i\sigma},\eta_{c_1i\sigma},\xi_{c_2i\sigma},\eta_{c_2i\sigma}\right)^T&
\end{align}

The computation details of the single-dot basis are presented in App.~\ref{App:ReviewCOM}.

\paragraph{Dimer basis}
The dimer-basis used is the following
\begin{align}
&\mathbf{\Psi}_{i\sigma} = \left(\mathbf{\Psi}_{\text{dimer},i\sigma},\xi_{c_2i\sigma},\eta_{c_2i\sigma}\right)^T&
\end{align}
with $\mathbf{\Psi}_{\text{dimer}}$ dimer of $f$ and $c_1$ electrons. \\

The computational details of the dimer basis are rather lengthy. We therefore provide additional files containing the expressions of the $I$ and $M$ matrices in terms of electronic operators (see the attached supplementary files).

\subsection{Lattice model benchmark\label{Sec:Benchmark_DQMC}}

In this work, the THFM at charge neutrality is investigated using ALF which is free of the fermion sign problem, enabling simulations at inverse temperatures $\beta = 100$ and for system sizes up to $L_x = 100$. The imaginary-time Green’s functions obtained from the Monte Carlo simulations can be analytically continued to real-frequency spectra using the stochastic and Bayesian analytic continuation procedures implemented in the ALF package~\cite{Beach,SandvikPRB,SHAO20231,Nevanlinna,JarrellBaye,vonderLinden}. However, analytic continuation is an ill-conditioned problem. Since the HO method provides an analytical expression for the GF, it can be evaluated directly in imaginary time and compared quantitatively with the DQMC data, providing an benchmark for the imaginary time GF.

\begin{figure}[h!]
    \centering
    \includegraphics[width=1.0\linewidth]{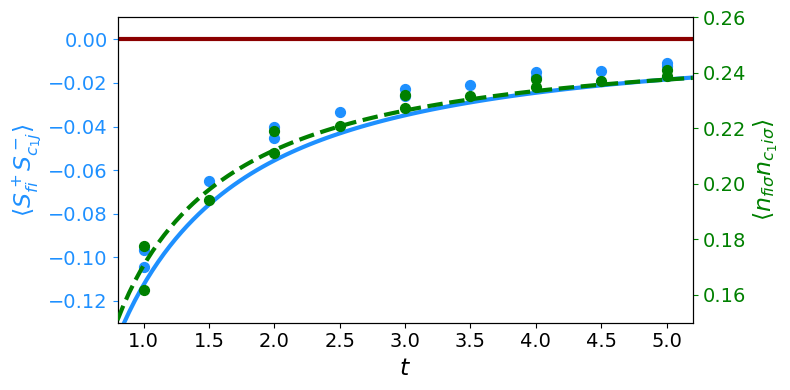}
    \caption{Local correlations $\langle S^+_{fi} S^-_{c_1 i} \rangle$ (blue) and $\langle n_{fi\sigma} n_{c_1 i\sigma} \rangle$ (green) are shown for the lattice model (Eq.~\ref{eq:THFM1_model}) with $U=1.5$, $\gamma=1.0$ as a function of hopping $t$. COM (lines) agrees with DQMC (dots) from small to large $t$, while Hubbard-I (red) fails away from the very large $t$ limit.}
    \label{fig:fig3}
\end{figure} 

\subsubsection{Static correlation functions from Roth decoupling}
In Fig.~\ref{fig:fig3}, we present correlation functions that quantify whether the $f$ electrons behave as local moments in the lattice-regularized topological heavy-fermion model and their coupling with the itinerant electrons. In Fig.~\ref{fig:fig3} we show $\langle S^+_{fi} S^-_{c_1i} \rangle$, $\langle n_{fi} n_{c_1i} \rangle$ as a function of hopping parameter $t$, such that the model is tuned toward the local moment regime where $W_{2} \gg U,\gamma$. COM and DQMC results differs at most by $10\%$ and exhibit a crossover toward the Hubbard-I limit where the $f$ and $c_1$ correlation functions are decoupled.

Furthermore, we analyze the temperature dependence of the correlation functions of Hubbard-I and COM. For the THFM model, we compare COM results with numerically exact DQMC data in Fig.~\ref{fig:Appfig3} as a function of temperature $T$, focusing on the same correlations. The COM results follow the DQMC trend across the entire temperature range. At low temperature, the correlation functions remain significantly different from their Hubbard-I values, indicating that the Hubbard-I approximation is a high temperature approximation that cannot capture the relevant physics of the local-moment regime in MATBG if non-local correlations are important.

From this, we can identify two distinct regimes: a correlated regime and a trivial one. In the former, significant Kondo hybridization occurs between $f$ and $c_1$, whereas in the trivial regime the two degrees of freedom are effectively decoupled. In the latter regime, since local moments are effectively decoupled, one therefore expects the COM and Hubbard-I approximations to accurately capture the spectral properties in the trivial regime. However, this picture breaks down in the regime where $U \ge \gamma$, even when $W_{2} \gg U,\gamma$. In this case, the gap is strongly renormalized and can be reduced to values as small as $U/3$, in agreement with DQMC results, instead of $U$ predicted by Hubbard-I and COM. Both these approximations treat the $f$-$c_1$ and $c_1$-$c_2$ hybridizations independently. This suggests that such a description is insufficient and that the influence of the $c_2$ electrons on the resulting correlation functions must be treated more carefully, which is achieved within the dimer basis. Even when Kondo hybridization is weak, the local-moment regimes have short range correlations and cannot be adequately captured within a purely local description. 

\subsubsection{Single-particle energy gap}
The benchmark of the imaginary-time spectral function from our different basis against DQMC can provide an estimate of real frequency  without having to perform analytic continuation. The spectral function in real frequency $A(\omega)$ is related to the imaginary time Green's function as

\begin{align}
& \mathcal{G}(\tau) = -\int d\omega \dfrac{e^{-\tau\omega}}{1+e^{-\beta\omega}}A(\omega)&
\end{align}
with $\tau \in [0,\beta]$. At larger $\tau \sim \beta/2$, contributions from energies $\omega \gtrsim T$ are exponentially suppressed, making the Green’s function primarily sensitive to low-energy physics. For a system with gap $\Delta$, one expects $G(\beta/2) \sim e^{-\beta \Delta/2}$ to be strongly suppressed. By studying $G(\beta/2)$, one can then infer the amount of low-frequency spectral weight.

\subsubsection{Imaginary time Greens function }
\begin{figure}
    \centering
\includegraphics[width=1.0\linewidth]{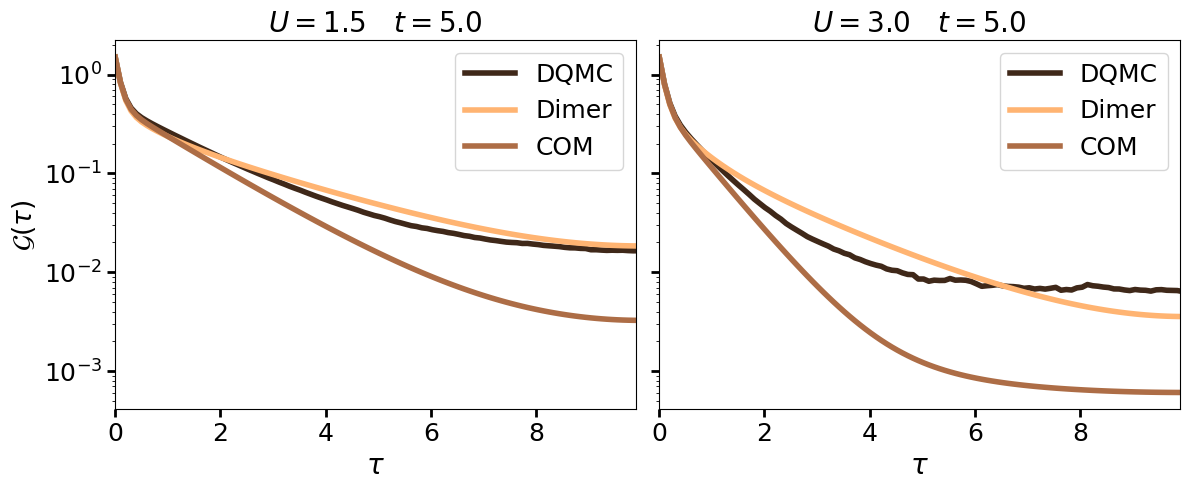}
    \caption{$\mathcal{G}(\tau,k=\pi/2)$ obtaind with COM, dimer approximations and unbias DQMC results. The parameters are $U=1.5$, $t=3.0$ (left panel) and $U=3.0$, $t=5.0$ (right panel)}
    \label{fig:fig4}
\end{figure}

In Fig.~\ref{fig:fig4}, we benchmark the imaginary-time Green’s function against DQMC results at $k=\pi/2$ where the local-moment regime is expected to be most pronounced. The benchmarks are performed in a parameter regime satisfying $W_{2} \gg U$. The COM approximation systematically underestimates the Green’s function compared to DQMC, indicating a incorrect stronger tendency toward a gapped, local-moment regime. For $U=3.0$ and $t=5.0$, the discrepancy exceeds one order of magnitude, suggesting a significant overestimation of the gap that cannot be explained by spectral-weight redistribution at higher energy. In contrast, the dimer HO results follows the DQMC trend over the entire $\tau$ range. Some discrepancies are still present, which is not surprising since the basis contains insufficient information to produce the exact results. Nevertheless, dimer HO already constitutes a significant improvement over the COM $\mathcal{G}(\tau)$. 

To quantify the deviation of the imaginary-time Green’s function obtained with  approximate calculations and that from the exact DQMC data  we compute the following momentum dependent figure of merit $\Delta \mathcal{G}(k)$ 
\begin{align}
&\Delta \mathcal{G}(k)\equiv \sum_\tau |\mathcal{G}_{HO}(\tau,k)-\mathcal{G}_{DQMC}(\tau,k)|.&
\end{align}

\begin{figure}[h!]
    \centering
\includegraphics[width=1.05\linewidth]{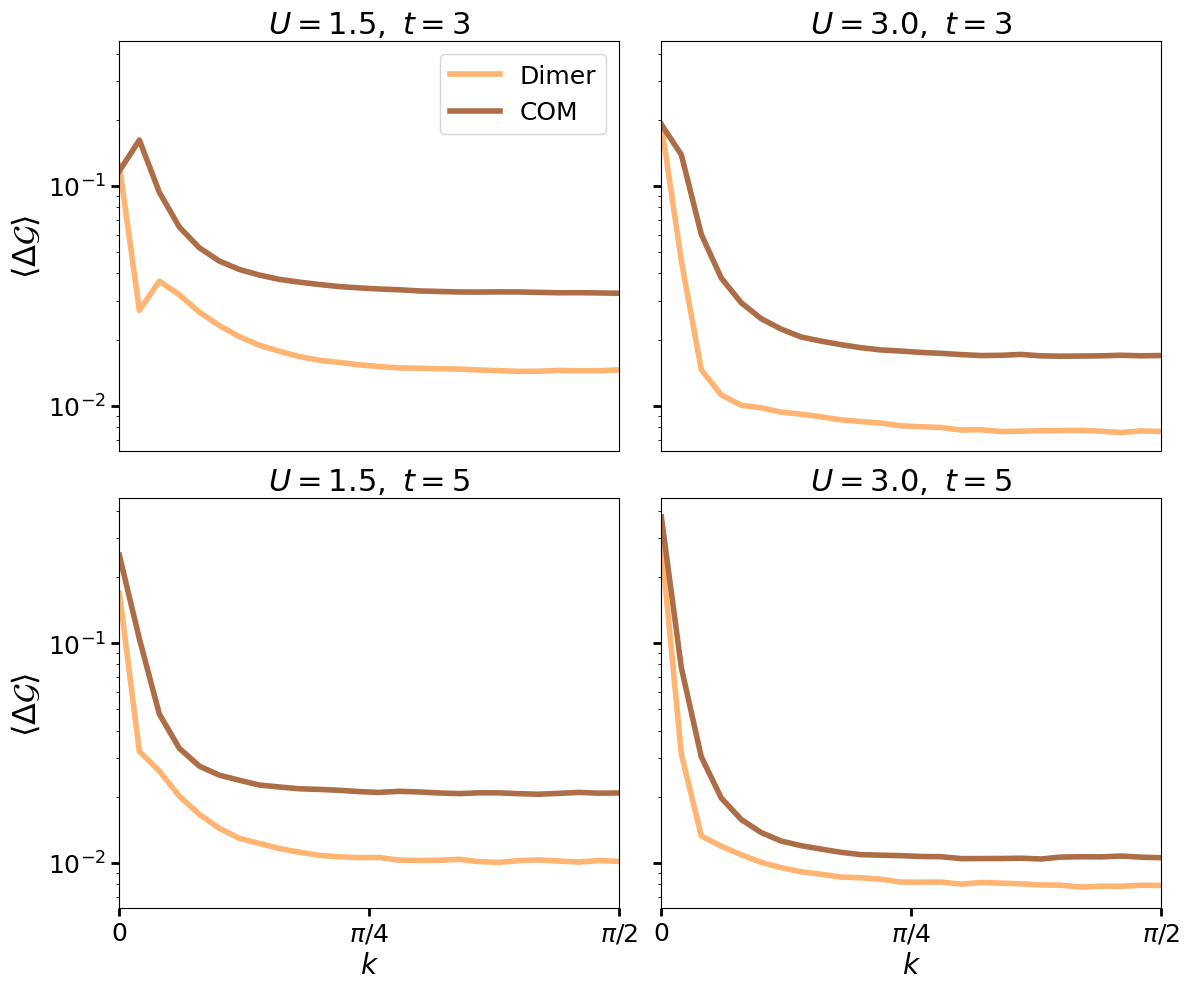}
    \caption{$\Delta \mathcal{G}(k)$ as a function of momentum for the COM and and dimer approximations. The parameters are $U=1.5$ (left panel) and $U=3.0$ (right panel). The upper panels correspond to $t=3$, while the lower ones correspond to $t=5$.}
    \label{fig:fig5}
\end{figure}

We present the results for two different values of the $c_1$--$c_2$ hopping $t$ in Fig.~\ref{fig:fig5}. In all cases, the dimer basis is in better agreement with the DQMC results than the single-site basis, sometimes significantly. This first highlights that the correlation functions included in the single-site basis are insufficient to fully capture the physics of the model for realistic parameters. More generally, it demonstrates the usefulness of the HO method as a systematic approximation scheme, whose accuracy can be improved by incorporating an increasingly larger set of correlation functions.

\section{Spectral function\label{Sec:Results}}
In this section, we describe the spectrum of the THFM. The quasiparticle broadening used in the COM, Hubbard-I, and dimer-basis approximation are set to $\varepsilon= 0.025$. The static correlations for the dimer calculations are obtained from DQMC simulations of $L=64$, with trotter time step $d\tau = 0.005$, and inverse temperature fixed to $\beta = 20$.

Because we study a lattice-regularized model the large-momentum limit considered in continuum models is best approximated around $k=\pi/2$. This will be used to diagnose the local moment behavior of the exact DQMC and HO calculations.

We begin by comparing DQMC analytic continuation results with HO approximations to provide insight into the real-frequency spectral properties. We choose parameters in the regime $W_{2} \sim \gamma \sim U$, to access a reliable analytic continuation and the DQMC spectral functions are computed for $L=100$.

\begin{figure}[h!]
    \includegraphics[width=1.\linewidth]{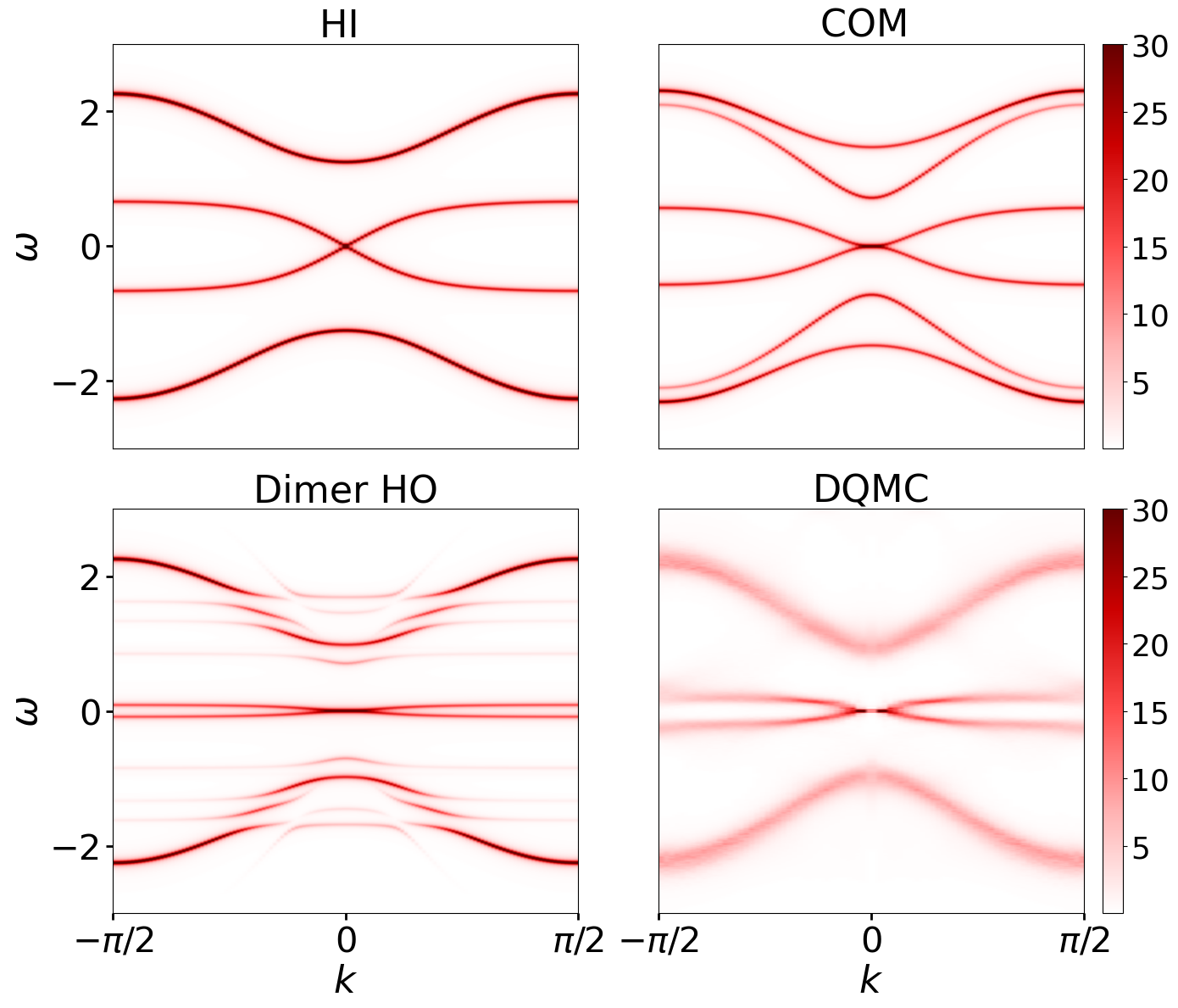}
    \caption{Spectral function of the  THFM. The top panel shows the Hubbard-I (left) and COM (right) results, and the bottom panel the dimer (left) and DQMC (right) results. The parameters are $U =1.5$, $t=1.0$, $\gamma=1.0$ and $T= 0.05$.}
    \label{fig:fig6}
\end{figure}
In Fig.~\ref{fig:fig6}, we compare the electronic spectra obtained using the HO method for single-dot and dimer basis and DQMC simulations. For finite $U$, the low-energy $f$-electron flat bands split as a result of local-moment formation. As the momentum approaches the $\Gamma$ point, the $f$-electron spectral weight in these flat bands is progressively suppressed, leading to a closing of the gap at $k = 0$ and the emergence of a Mott semimetal phase. This behavior can be traced back to the orbital composition of the flat band, which involves a hybridization between $f$ and $c_2$ orbitals. Consequently, low-energy excitations near the $\Gamma$ point acquire predominantly $c_2$ character. In contrast, the two higher-energy dispersive bands, inherited from the noninteracting limit, retain their mixed $f$, $c_1$, and $c_2$ orbital character.

The band crossing at the $\Gamma$ point is not linearly dispersing, in contrast to the projected limit~\cite{hu2025projected,cualuguaru2025obtaining,ledwith2025exotic,ledwith2025nonlocal}, where $U \ll \gamma$. This deviation originates from the coupling of conduction electrons to  local moments, as well as from the inclusion of nonlocal correlations. In the high-temperature limit, COM reduces to the Hubbard-I approximation. In that regime, where such scattering processes are suppressed, the crossing at the $\Gamma$ point recovers a linear dispersion.

The key physical signature of the local-moment regime is the spectral gap in the spectral function near $k=\pi/2$. For the Hubbard-I approximation, the single-particle gap is $U$, as expected since it describes a decoupled local moment. However, COM captures some correlations and the gap is slightly smaller than $U$. For the dimer and DQMC results, the gap is significantly smaller than $U$, indicating that the picture of decoupled local moments is not applicable in this regime. Therefore the single-site basis naturally favors local-moment description, although this tendency is partially reduced in the COM where the Roth decoupling introduces certain non-local correlations. A similar limitation arises in embedding-based local approaches such as DMFT~\cite{cualuguaru2025obtaining}.

\subsection{Energy gap from imaginary-time data}
\begin{figure}[h!]
    \centering
    \includegraphics[width=1.05\linewidth]{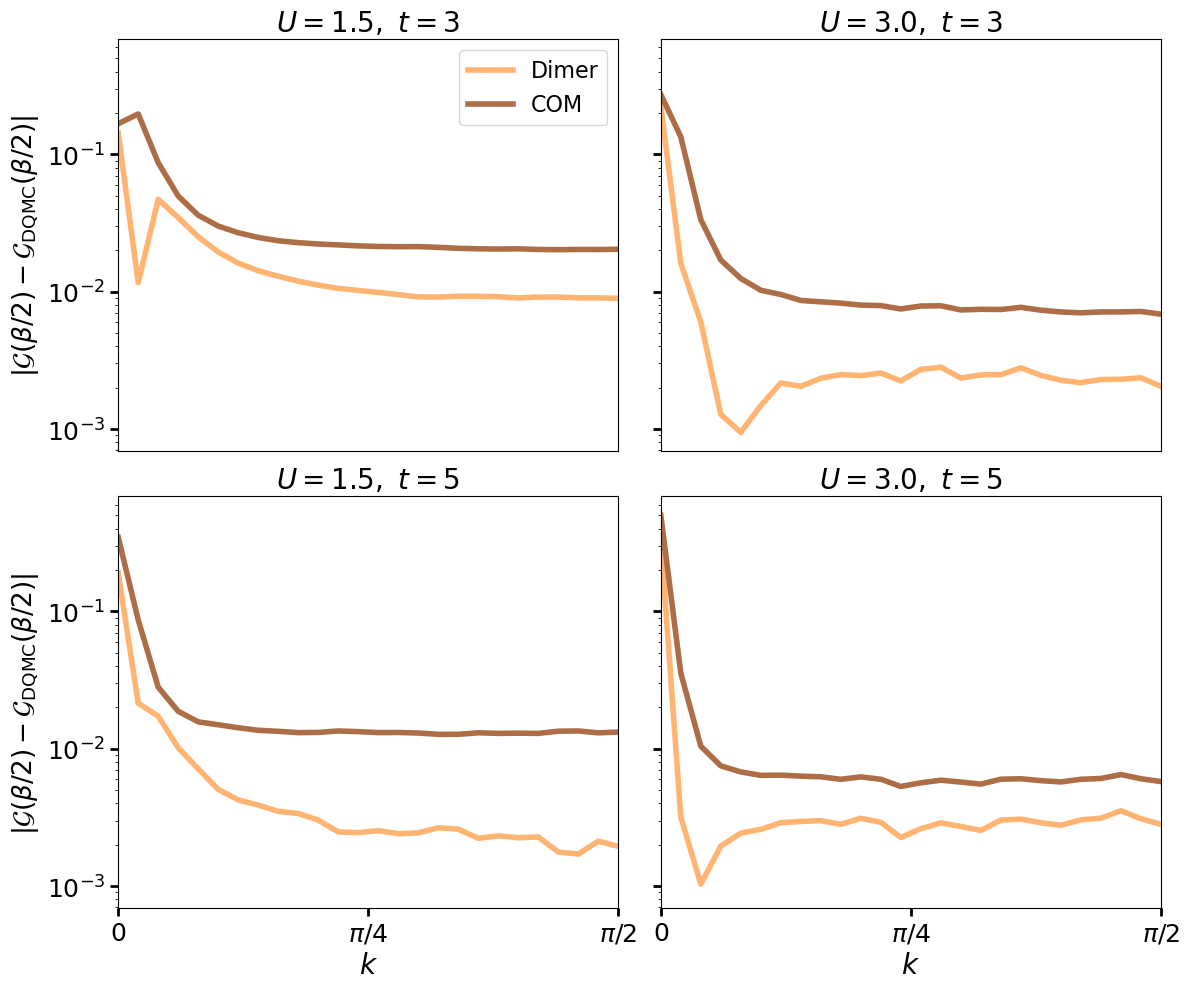}
    \caption{ $|\mathcal{G}(\beta/2,k)-\mathcal{G}_{DQMC}(\beta/2,k)|$ as a function of momentum for the COM, and dimer approximations. The parameters are $U=1.5$ (left panel) and $U=3.0$ (right panel). The upper panels correspond to $t=3$, while the lower ones correspond to $t=5$.}
    \label{fig:fig7}
\end{figure}

We study $\mathcal{G}(\tau=\beta/2)$  to provide an estimate of the gap without relying on the ill-conditioned analytic continuation procedure for DQMC. As discussed in Sec.~\ref{Sec:Benchmark_DQMC} $\mathcal{G}(\beta/2,k)$ can be used to probe the spectral function gap as a function of momentum. In Fig.~\ref{fig:fig7}, this quantity is shown for the HO method and compared with DQMC results for $U \sim 1.5\gamma$ and $U \sim 3\gamma$ in the regime $W_{2} \gg U,\gamma$ as a function of momentum.

In Fig.~\ref{fig:fig7} COM captures the free local-moment regime consisting of decoupled Hubbard dots~\cite{hu2025projected}. However, the results show that COM approximation substantially overestimate the gap over a wide momentum range. In contrast, near $k=\pi/2$, the dimer approximation reproduces the DQMC gap more accurately.

In general, the dimer basis yields a gap that remains significantly closer to the DQMC results across both interaction regimes. As shown in Fig.~\ref{fig:fig4}, it tends to underestimate the single-particle gap, reflecting the bias toward Kondo coupling between the $f$ and $c_1$ electrons at the expense of the coupling between the $c_1$ and $c_2$ electrons.

The physical origin of this behavior can be understood from the simple dimer Hubbard model discussed in App.~\ref{Sec:Benchmark_Dimer}. A key point in this regime is that $U$ must be much larger than $\gamma$  to generate well-defined Hubbard peaks. In the experimentally relevant regime $U \sim (1\text{--}5)\gamma$, this condition is not fulfilled: the gap becomes strongly renormalized and the spectral features are substantially broadened, which is shown with DQMC. This shows that non-local correlations are important to predict the correct local-moment physics in the lattice regularized THFM model. 

Although the dimer approximation is capturing better the gap, the accuracy is reduced for smaller imaginary time $\tau$ as shown in Fig.~\ref{fig:fig4}. This indicates the distribution of spectral weight is still approximate. This is originating from the number of operators in the dimer basis, which is not large enough to capture finer details. 

Finally, the imaginary-time Green’s function imposes important constraints on the low-energy spectra obtained from DMFT, in particular regarding the existence of a quasiparticle peak, which is discussed in App.~\ref{App:QP_peak}.

\subsection{HO spectral function}

\begin{figure}[h!]
    \centering
    \includegraphics[width=1.0\linewidth]{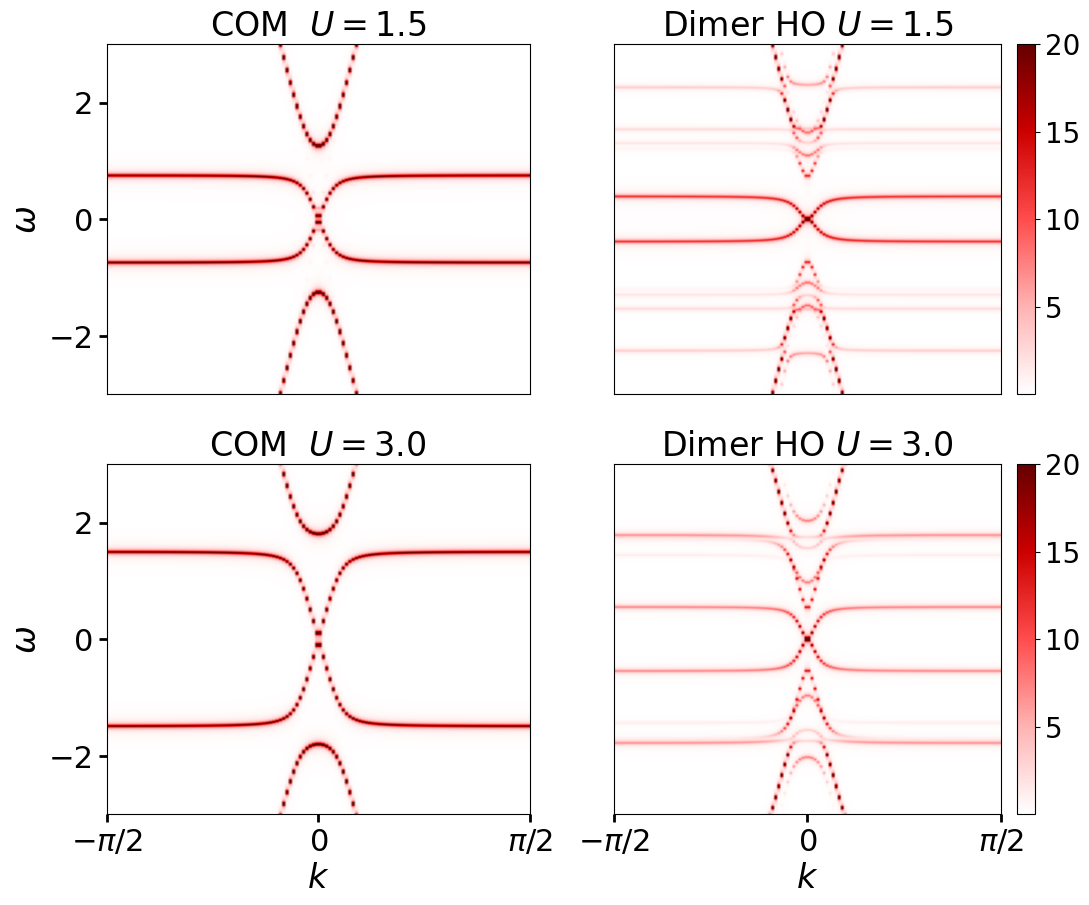}
    \caption{Spectral function of the  THFM. The left panel shows the COM results, and the right panel the dimer results. The parameters are $t=5.0$, $\gamma=1.0$ and $T= 0.05$. The top panels are for $U =1.5$  and the bottom ones are for $U=3.0$} 
    \label{fig:fig9}
\end{figure}
We finally present the Hubbard-operator spectral function in real frequency for both the single-site and dimer basis in the regime $W_{2} \gg \gamma, U$~\cite{cualuguaru2025obtaining}. Fig.~\ref{fig:fig9} shows the results for fixed $U=1.5$ as a function of $t$.

As discussed previously, COM effectively decouples the $f$ electron away from the $\Gamma$ point, producing a gap of order $U$. This behavior highlights the importance of non-local correlation functions for reproducing the exact spectral features. In contrast, the gap obtained within the dimer approximation remains substantially smaller than $U$. For fixed $t$ and $\gamma$, the spectral weight of the lower-energy branch decreases with increasing $U$ and is transferred to higher-energy excitations. This redistribution indicates that the electronic self-energy grows with interaction strength, consistent with previous studies~\cite{hu2026twistedbilayergraphenelifetimes,vituri2026controlledloopexpansiontopological,wei2026lifetimespectralfunctiontopological}. In the present case, this effect arises from nonperturbative correlations.

In Fig.~\ref{fig:fig9} bottom panels, we present the spectrum at large $t=5$ and larger interactions strength $U=3$. In this limit, one would expect the system to approach the local-moment regime. Even in this regime, where charge fluctuations are strongly suppressed, the single-particle gap is significantly reduced in the dimer basis compared to the single-dot basis.

However, the dimer basis remains too limited to accurately capture the imaginary part of the self-energy. As the composite-operator basis is enlarged, the sharp low-energy band is expected to broaden into a continuum of spectral weight. In the range $U \sim (1\text{-}3)\gamma$, the dimer gap remains approximately $\Delta \sim U/3$, indicating that the system is still far from the local-moment limit. The contrast between the COM and dimer results underscores the importance of treating nonlocal correlations accurately.

\section{Conclusion}
We apply the HO method to the topological heavy-fermion model beyond both the fully decoupled local-moment limit and the projected-band regime. The single-site basis, equivalent to COM at lowest order, captures static local correlation effects but remains insufficient to describe the full spectral features. COM reproduces the Hubbard-I limit in the absence of intersite correlations but deviates once nonlocal processes become relevant. In contrast, the exact spectrum is recovered only when intercluster correlations are included beyond the local limit.

The spectral gap away from the Dirac point is significantly reduced compared to local approximations, demonstrating the importance of nonlocal correlations in the model. This effect persists even for $N_f=2$, where correlation effects are expected to be weak, given that the self-energy away from the $\Gamma$ point scales approximately linearly with the flavor number, $\text{Im}\Sigma \propto N_f$~\cite{hu2026twistedbilayergraphenelifetimes,wei2026lifetimespectralfunctiontopological,vituri2026controlledloopexpansiontopological}. Here we find that nonlocal corrections are quantitatively relevant, leading to a large imaginary part of the self-energy even at low flavor number.

Different cluster choices emphasize different physical processes. The single-site basis favors local-moment physics, while the dimer basis incorporates short-range hybridization effects and is more consistent with Kondo-like behavior. However, the dimer basis still underestimates the gap and does not fully capture $c_1$-$c_2$ hybridization effects, which would require larger clusters such as a trimer basis. Such extensions are, however, computationally demanding within the present framework. Yet the HO approach provides a systematic and extendable hierarchy of approximations that goes beyond local self-energy constructions and allows to interpolate between local-moment and itinerant regimes.

The COM approximation is self-consistent due to the use of Roth decoupling, in contrast to the dimer basis, where the correlation functions are obtained from DQMC calculations. Within the COM framework, this leads to a self-consistent set of equations involving only four correlations. Despite this minimal number of correlation functions, the agreement with DQMC is quantitative across all model parameters and temperatures. The same Roth decoupling can be extended to a larger HO basis. The resulting self-consistent equations would involve a larger set of operators, making the approach significantly more involved. In contrast to other Green’s-function techniques such as DMFT, CPT, and related variants, the HO approach retains more microscopic information by explicitly tracking the properties of all charge-one excitations, rather than only those of the single-particle electronic excitations. A brute-force enlargement of the basis is limited and the approach cannot be practically implemented without introducing additional approximations.

\section{Acknowledgments}
The authors acknowledge F.~Assaad for providing the open-source ALF package and S.~Biswas for helpful discussions regarding the implementation of the THFM model in ALF. All numerical calculations were performed on the Kanta cluster at IPhT.
\appendix

\section{ Single-dot basis\label{App:ReviewCOM}}

\subsection{Topological heavy fermion model\label{App:TopoModel}}
In this section, we provide the equations for the single-site basis for the topological heavy fermion model with Hamiltonian:
\begin{align}
\nonumber\mathcal{H}_{\rm THFM} = &U\sum\limits_i \left(\hat{n}_{fi\uparrow}-1/2\right)\left( \hat{n}_{fi\downarrow}-1/2\right)&\\
\nonumber&-\gamma\sum\limits_{i\sigma}\left[f^\dagger_{i\sigma}c_{1i\sigma}+h.c.\right] - \sum\limits_{\langle ij\rangle\sigma}t_{ij}\left[c^\dagger_{1i\sigma} c_{2j\sigma}+h.c.\right]. &
\end{align}

The operator basis in the COM approximation reads
\begin{align}
&\Psi_{i\sigma} = \left(\xi_{fi\sigma},\eta_{fi\sigma},\xi_{c1i\sigma},\eta_{c1i\sigma},\xi_{c2i\sigma},\eta_{c2i\sigma}\right)^T&
\end{align}

In the paramagnetic limit, the I-matrix is purely local and diagonal
\begin{align}
I_{ij}=\delta_{ij}\text{diag}\left(1- n_{fi\overline{\sigma}}, n_{fi\overline{\sigma}}, 1-n_{c1i\overline{\sigma}}, n_{c1i\overline{\sigma}}, 1-n_{c2i\overline{\sigma}}, n_{c2i\overline{\sigma}} \right),
\end{align}
where $n_\alpha=\langle c^\dagger_{\alpha} c_{\alpha} \rangle$ is the local density. 
Assuming translation symmetry, the $M$ matrix can be expressed as
\begin{align}
&M(k) = \begin{pmatrix}
M_{ff} & M_{c_1f} & 0\\
M_{fc_1} & M_{c_1c_1} & M_{c_2c_1}\\
0 & M_{c_1c_2} & M_{c_2c_2} \\
\end{pmatrix}&\\
&M_{ff} = \begin{pmatrix}
(-U/2) \left(1-n_{f\overline{\sigma}}\right) & 0 \\
0 & (U/2) n_{f\overline{\sigma}} \\
\end{pmatrix}&\\
&M_{c_1f}=\begin{pmatrix}
-\gamma \left(1-n_{c1\overline{\sigma}}-n_{f\overline{\sigma}}+p_{cf}\right)& -\gamma\left(n_{c_1\overline{\sigma}}-p_{cf}\right)\\
-\gamma\left(n_{f\overline{\sigma}}-p_{cf}\right)& -\gamma p_{cf}\\
\end{pmatrix}&\\[5pt]
&M_{c_1c_2}=-2t\sin(k)\begin{pmatrix}
1-n_{c_1\overline{\sigma}}-n_{c_2\overline{\sigma}}+p_{c}& n_{c_1\overline{\sigma}}-p_{c}\\
n_{c_2\overline{\sigma}}-p_{c}& p_{c}\\
\end{pmatrix}&\\[5pt]
&M_{c_1c_1} =0_2\\
&M_{c_2c_2} = 0_2&
\end{align}

Here we define $S^{+}_{\alpha,i}=\alpha^{\dagger}_{i,\uparrow}\alpha_{i,\downarrow}, S_{\alpha,i}^{-}= (S^{+}_{\alpha,i})^{\dagger}$ and $\Delta_{\alpha,i}=\alpha_{i,\uparrow} \alpha_{i,\downarrow}$, with $\alpha\in\{f,c_1,c_2\}$. Then, the $p_{c}$ and $p_{cf}$ expectation values are written in terms of these as
\begin{align}
& p_{c} = \langle n_{c_1i\overline{\sigma}}n_{c_2i\overline{\sigma}}\rangle+\langle S_{c_1i}^+S_{c_2i}^ -\rangle -\langle \Delta^\dagger_{c_1i}\Delta_{c_2i}\rangle &\\
& p_{cf} = \langle n_{c_1i\overline{\sigma}}n_{fi\overline{\sigma}}\rangle+\langle S_{c_1i}^+S_{fi}^ -\rangle -\langle \Delta^\dagger_{c_1i}\Delta_{fi}\rangle, &
\end{align}
They are computed through Roth decoupling~\cite{roth1969electron, haurie2024bands,Pangburn_Layer,Pangburn_Ringstates,Pangburn_TMZ} in the COM approximation. The Hubbard I approximation consists of setting $p_{cf} = 0.25$ and $p_{c}=0.25$.

\subsection{Self-consistent scheme\label{App:SelfConsistency}}
The $M$ and $I$ matrices depend on certain expectation values that should be computed self-consistently. Some of these expectation values, such as the density operator $n_{i\sigma} = \langle c^\dagger_{i\sigma}c_{i\sigma} \rangle$, can be directly computed from the advanced and retarded composite Green's function. To do so, we define the correlation function matrix $\mathcal{C}$ of two operators in the local basis is computed as
\begin{align}
&\mathcal{C}_{\alpha \gamma}= -\int \frac{d\omega}{4 i \pi} \left[1+\tanh\left(\frac{\beta \omega}{2}\right) \right] \left(\mathds{G}^R(\omega))-\mathds{G}^A(\omega)\right)_{\alpha \gamma}.&\label{Eq:FullCorr}
\end{align}
For instance, $\langle n_{i\sigma}\rangle$ is expressed as 
\begin{align}
&\langle c^\dagger_{i\sigma} c_{i\sigma} \rangle = \langle \xi_{i\sigma}^\dagger\xi_{i\sigma}\rangle+\langle\xi_{i\sigma}^\dagger\eta_{i\sigma}\rangle+\langle\eta_{i\sigma}^\dagger\xi_{i\sigma}\rangle+\langle\eta_{i\sigma}^\dagger\eta_{i\sigma}\rangle &
\end{align}
Other expectation values, such as $\langle n_{i\uparrow} n_{j\uparrow} \rangle$ that enter in the $p$ parameter, cannot be computed directly and require an additional decoupling scheme. Roth decoupling~\cite{roth1969electron} is a possible scheme that relies on the operator algebra. The details of the decoupling procedure in Refs.~\cite{roth1969electron,Pangburn_Layer,haurie2024bands,banerjee2025charge,banerjee2025probing,Pangburn_Ringstates,Pangburn_TMZ}.

\subsection{Limitation of composite operator method\label{App:Limitation}}
We have performed the self-consistent calculation only for the lattice-regularized model. Nevertheless, the qualitative features of the spectrum near the $\Gamma$ point are governed by low-energy physics and are therefore insensitive to lattice details. The COM approach relies on an enlarged operator basis, allowing for couplings between composite operators via hopping processes. A natural extension beyond this local formulation would be to construct a decoupling scheme directly in momentum space, enabling a treatment of continuum models. Another approach would be to study longer range correlations as performed in Ref.~\cite{nosov2026controlled}. Extending the formalism beyond the $SU(2)$-symmetric case is challenging but remains feasible by enlarging the operator basis. The $SU(4)$ case can be treated within a dimer basis, while the $SU(8)$ case relevant to magic-angle twisted bilayer graphene (MATBG) can be analyzed directly using a quadrimer basis.

\section{Benchmark dimer problem\label{Sec:Benchmark_Dimer}}
In this section, we benchmark the single-site HO method in the Anderson dimer limit~\cite{Correl21}.

\subsection{Dimer limit\label{Sec:DimerLimit}}
At $k=0$, the equations of motion for the gapless THFM nearly close: the $c_2$ electrons are decoupled, and the residual $c_1-c_2$ coupling enters only at higher order. The system can be effectively treated as a $c_1-f$ dimer and the spectra at $k=0$ can be understood via the dimer Hubbard model
\begin{align}
\mathcal{H}_{\rm d}= &U\left(f^\dagger_{\uparrow}f_{\uparrow}-1/2\right)\left(f^\dagger_{\downarrow}f_{\downarrow}-1/2\right)&\\
&-\gamma \sum\limits_{\sigma} \left[ f_{\sigma}^\dagger c_{\sigma}+h.c \right]&
\label{eq:Dimer_Hamiltonian}
\end{align}
$U$ and $\gamma$ are the on-site Hubbard repulsion and the hybridization amplitude respectively.

The single-site operator basis for the dimer problem read
\begin{align}
&\Psi_{\sigma} = \left( \xi_{f\sigma},\eta_{f\sigma},\xi_{c\sigma},\eta_{c\sigma} \right)^T&
\end{align}

where $\xi_{c\sigma}$ and $\eta_{c\sigma}$ are holons and doublons for the $c$-orbital. At half filling, the $M$-matrix decomposes into $2\times 2$ blocks:
\begin{align}
&M =
\begin{pmatrix}
M_{ff}& M_{fc} \\
M_{cf} & M_{cc} \\
\end{pmatrix}
\end{align}
with
\begin{align}
&M_{ff} =\begin{pmatrix}
-U/2\left(1-n_{f\overline{\sigma}}\right)& 0 \\
& \\
0& U/2n_{f\overline{\sigma}} \\
\end{pmatrix}&\\[5pt]
&M_{fc} = \begin{pmatrix}
 -\gamma\left(1-n_{f\overline{\sigma}}-n_{c\overline{\sigma}}+p_{cf}\right)& -\gamma\left(n_{f\overline{\sigma}}-p_{cf}\right) \\
-\gamma\left(n_{c\overline{\sigma}}-p_{cf}\right) & -\gamma p_{cf} \\
\end{pmatrix}&\\[5pt]
&M_{cf} = M_{fc}^\dagger&\\
&M_{cc} =0&
\end{align}

The diagonal normalization matrix is
\begin{align}
&I= \text{diag} \begin{pmatrix}
1-n_{f\overline{\sigma}}, n_{f\overline{\sigma}}, 1-n_{c\overline{\sigma}},  n_{c\overline{\sigma}} \\
\end{pmatrix}&
\end{align}
The additional parameter $p_{cf}$ captures density–density, spin–spin, and pair–pair correlations:
\begin{align}
&p_{cf} = \langle n_{c\overline{\sigma}}n_{f\overline{\sigma}} \rangle +\langle S^+_{c}S_f^-\rangle -\langle \Delta^\dagger_{c}\Delta_f\rangle&
\label{eq:p_parameter}
\end{align}
and is determined self-consistently through Roth decoupling.

\subsection{Zero-temperature limit}
\begin{figure}[h!]
    \centering
    \includegraphics[width=1.\linewidth]{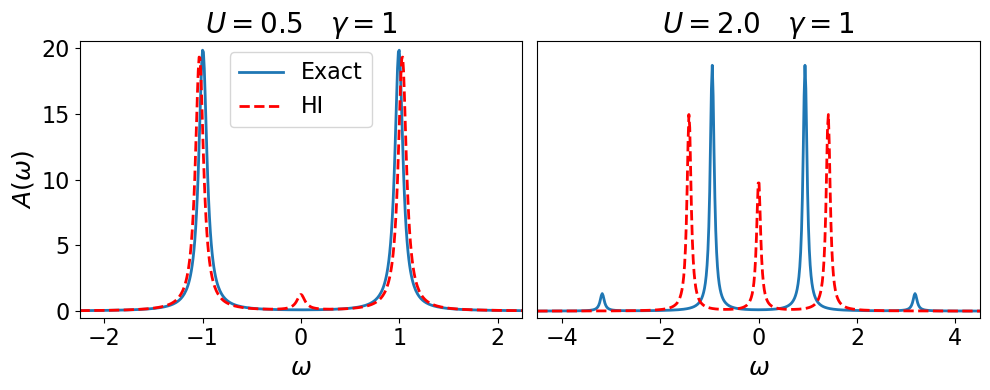}
    \caption{Comparison of the spectral function $A(\omega)$ for the dimer problem between the exact solution (blue curve) and the Hubbard-I approximation (red dashed curve), shown for $U=0.5$ in the left panel and $U=2.0$ (right panel) at $\gamma=1.0$ for $T=0$.}
    \label{fig:Appfig1}
\end{figure}
For the dimer at half-filling at $T=0$, the Roth decoupling  becomes exact and therefore COM reproduce the exact spectrum for all $U$. Fig.~\ref{fig:Appfig1} compares the exact spectral function at $T=0$ with the Hubbard-I approximation for several values of $U$ at fixed hybridization $\gamma = 1.0$. The Hubbard-I approximation, which restricts the Green’s function to a three-pole structure, fails to capture the exact spectrum, which contains four excitation peaks. This discrepancy reflects its treatment of nonlocal correlations as high-temperature decoupling, effectively disconnecting the two sites of the dimer for $U \neq 0$.

By contrast, the COM treats the $c-f$ interaction exactly at $T=0$ for the dimer at half-filling, capturing local correlations and reproducing the exact spectrum for all $U/\gamma$. Therefore, the COM provides the low-energy excitations to capture the physics of Anderson lattice systems at low temperatures.

\subsection{Finite temperature}
\begin{figure}[h!]
    \centering
    \includegraphics[width=1.\linewidth]{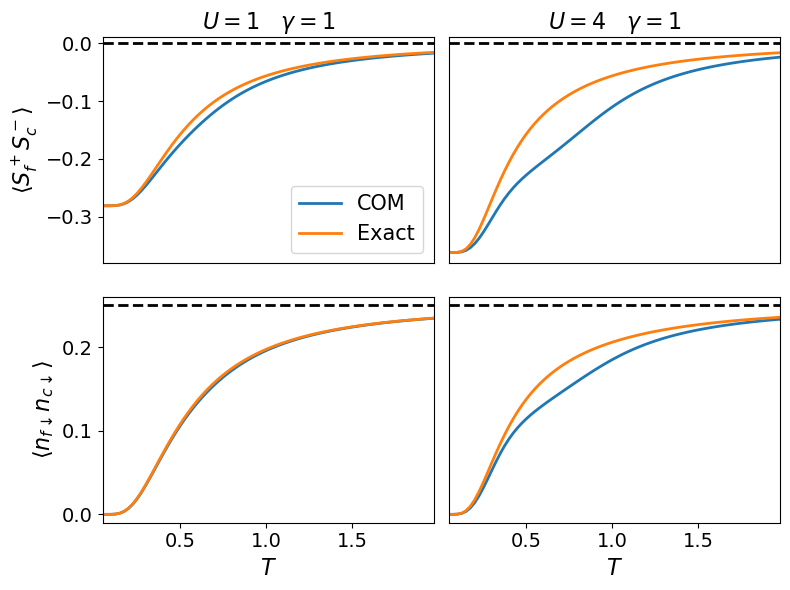}
    \caption{Temperature dependence of $\langle S^{+}_{f} S^{-}_{c}\rangle$ (top) and $\langle n_{f\downarrow}n_{c\downarrow}\rangle$ (bottom) for the dimer. Exact results are in orange, COM  in blue, and Hubbard-I in black are shown for $U=1.0$ (left) and $U=4.0$ (right), with $\gamma=1.0$.}
   
    \label{fig:Appfig4}
\end{figure}
For $T \neq 0$, the COM becomes approximate, as the four-pole ansatz no longer captures all thermally activated excitations. In Fig.~\ref{fig:Appfig4}, we benchmark inter-site correlation functions against the exact solution as a function of temperature. Despite its approximate nature, COM accurately reproduces the overall behavior across both weakly and strongly correlated regimes.

At low temperature, the COM approximation becomes exact because the Roth decoupling constraints are satisfied exactly. 

At higher temperatures, additional excitations beyond holons and doublons become relevant, and the COM loses accuracy. In the limit $T \to \infty$, COM reduces to the Hubbard-I approximation, where nonlocal correlations are fully decoupled. In this regime, Hubbard-I predicts a spurious zero-energy peak absent in the exact solution, highlighting its failure to describe the correct spectral properties at high temperature in both lattice and continuum models.

\begin{figure}[h!]
\centering
\includegraphics[width=1.\linewidth]{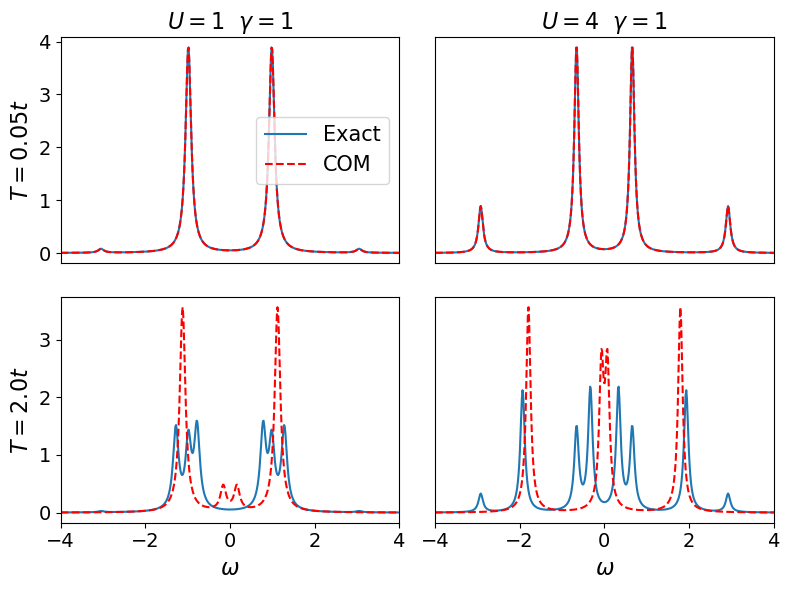}
\caption{Spectral function $A(\omega)$ for the dimer problem for exact (blue solid curve) and COM (red dashed curve), shown for  $T=0.05t$ in the top panels and $T=2.0t$ in the bottom panels. (Left) $U/\gamma=1.0$ (Right) $U/\gamma=4.0$}
\label{fig:fig13}
\end{figure}
In Fig.~\ref{fig:fig13}, the spectral functions from COM are compared with the exact solution as a function of temperature. At high temperature within the COM approximation, excitations associated with the operator basis become effectively decoupled on different sites, leading to the appearance of a zero-energy mode. This contrasts with the exact spectrum, which displays multiple peaks near zero energy together with significant spectral broadening. In the high-temperature limit, all many-body states become equiprobable as the correlations are suppressed and the spectral features becomes incoherent. Such incoherent features are not captured within the COM framework.

This comparison highlights a key limitation of COM: while it provides a reliable and self-consistent description of static correlation functions, its treatment of dynamical quantities is crude. This limitation is generic to equation-of-motion approaches, whose lowest-order truncations resemble mean-field theories. As a result, spectral functions obtained within COM should be interpreted with care, particularly at high temperatures.


\section{Additional results and benchmarks\label{Sec:AdditionalBenchmark}}

In this section, we present additional results supporting the discussion in the main text.

\subsection{Correlation function as a function of temperature\label{App:Corr_Temperature}}
\begin{figure}[h!]
    \centering
    \includegraphics[width=1.0\linewidth]{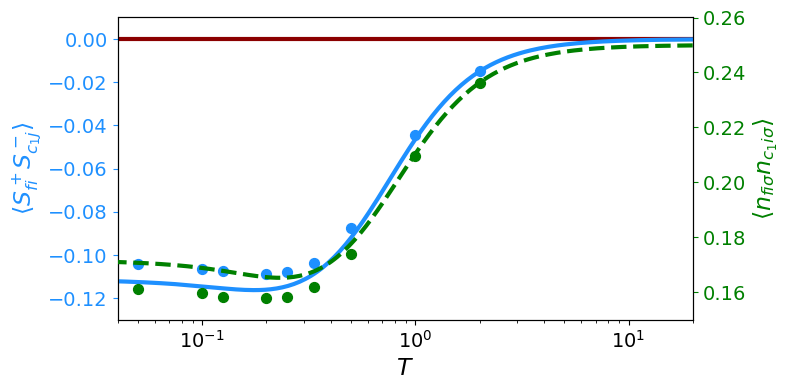}
    \caption{Local correlations $\langle S^+_{fi} S^-_{c_1 i} \rangle$ (blue) and $\langle n_{fi\sigma} n_{c_1 i\sigma} \rangle$ (green) are shown for the lattice model (Eq.~\ref{eq:THFM1_model}) with $U=1.5$, $\gamma=1.0$ and $t=1.0$ as a function of temperature. COM (lines) agrees with DQMC (dots) across temperatures, while Hubbard-I (red) fails away from the high temperature limit.}
    \label{fig:Appfig3}
\end{figure}

In Fig.~\ref{fig:Appfig3}, we present local correlation function between $f$ and $c_1$ electrons as a function of temperature. The main insight is that COM results are very close from DQMC data for the whole temperature range, from $T\ll t,U$ to $T\gg U$ where the system is completely incoherent. As the temperature increase, each degree of freedom completely decouple and approach the Hubbard-I limit.

\subsection{Quasiparticle peak bound\label{App:QP_peak}}

A quasiparticle peak contribution to the Green's function can be modeled as
\begin{align}
&G_{\mathrm{QP}}(\omega) = \frac{Z_{\mathrm{QP}}}{\omega + i\delta_{\mathrm{QP}}},&
\end{align}
where $\delta_{\mathrm{QP}}$ corresponds to the inverse quasiparticle lifetime of the QP peak. In DMFT the Kondo is sharp~\cite{cualuguaru2025obtaining} such that one can set to a good approximation $\delta_{\mathrm{QP}}=0$, and obtains
\begin{align}
&\mathcal{G}(\beta/2)=\mathcal{G}_{\mathrm{QP}}+\mathcal{G}_{\mathrm{others}},&
\end{align}
with $\mathcal{G}_{\mathrm{QP}}=\frac{Z_{\mathrm{QP}}}{2}$
and $\mathcal{G}_{\mathrm{others}}$ the contribution of the others Green's function poles.

\begin{figure}[h!]
    \centering
    \includegraphics[width=0.9\linewidth]{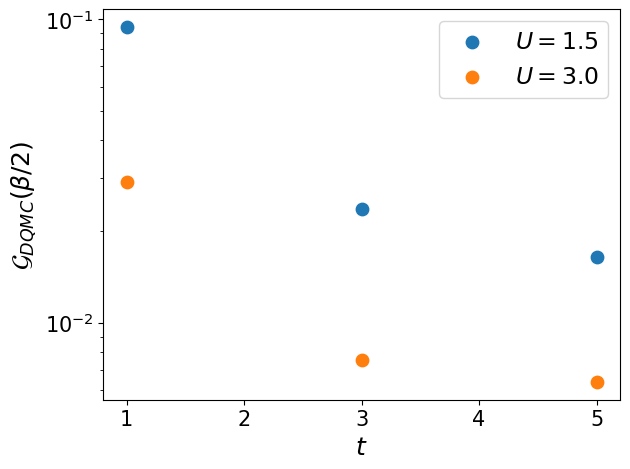}
    \caption{$\mathcal{G}_{\mathrm{DQMC}}(\beta/2)$ as a function of the hopping $t$ for $\gamma = 1$. The blue dots correspond to $U = 1.5$, while the orange dots correspond to $U = 3.0$.}
    \label{fig:fig8}
\end{figure}

This provides the bound $Z_{\mathrm{QP}}\leq2\,\mathcal{G}_{\mathrm{DQMC}}(\beta/2)$, which provides a constraint on the quasiparticle-peak spectral weight. The constraint is in practice stronger as $\mathcal{G}_{\mathrm{others}}$ can be large. The low values of $\mathcal{G}_{\mathrm{DQMC}}(\beta/2)$ shown in Fig.~\ref{fig:fig8} therefore disfavor the existence of a quasiparticle peak at low temperature. This is consistent with the HO method, irrespective of the basis size.

\bibliography{THFM_biblio.bib}

@article{hubbard1963electron,
  title={Electron correlations in narrow energy bands},
  author={Hubbard, John},
  journal={Proceedings of the Royal Society of London. Series A. Mathematical and Physical Sciences},
  volume={276},
  number={1365},
  pages={238--257},
  year={1963},
  publisher={The Royal Society London}
}

@misc{Beach,
      title={Identifying the maximum entropy method as a special limit of stochastic analytic continuation}, 
      author={K. S. D. Beach},
      year={2004},
      eprint={cond-mat/0403055},
      archivePrefix={arXiv},
      primaryClass={cond-mat.str-el},
      url={https://arxiv.org/abs/cond-mat/0403055}, 
}

@Article{ALFcode,
	title={{The ALF (Algorithms for Lattice Fermions) project release 2.4. Documentation for the auxiliary-field quantum Monte Carlo code}},
	author={F. F. Assaad and M. Bercx and F. Goth and A. Götz and J. S. Hofmann and E. Huffman and Z. Liu and F. Parisen Toldin and J. S. E. Portela and J. Schwab},
	journal={SciPost Phys. Codebases},
	pages={1-v2.4},
	year={2025},
	publisher={SciPost},
	doi={10.21468/SciPostPhysCodeb.1-v2.4},
	url={https://scipost.org/10.21468/SciPostPhysCodeb.1-v2.4},
}

@article{hubbard1964electron,
    author = {Hubbard, J.},
    title = {Electron correlations in narrow energy bands. II. The degenerate band case},
    journal = {Proceedings of the Royal Society of London. A. Mathematical and Physical Sciences},
    volume = {277},
    number = {1369},
    pages = {237-259},
    year = {1964},
    month = {01},
    issn = {0080-4630},
    doi = {10.1098/rspa.1964.0019}
}

@article{harris1967single,
  title = {Single-Particle Excitations in Narrow Energy Bands},
  author = {Harris, A. Brooks and Lange, Robert V.},
  journal = {Phys. Rev.},
  volume = {157},
  issue = {2},
  pages = {295--314},
  numpages = {0},
  year = {1967},
  month = {May},
  publisher = {American Physical Society},
  doi = {10.1103/PhysRev.157.295},
  url = {https://link.aps.org/doi/10.1103/PhysRev.157.295}
}

@article{roth1969electron,
  title = {Electron Correlation in Narrow Energy Bands. I. The Two-Pole Approximation in a Narrow $S$ Band},
  author = {Roth, Laura M.},
  journal = {Phys. Rev.},
  volume = {184},
  issue = {2},
  pages = {451--459},
  numpages = {0},
  year = {1969},
  month = {Aug},
  publisher = {American Physical Society},
  doi = {10.1103/PhysRev.184.451},
  url = {https://link.aps.org/doi/10.1103/PhysRev.184.451}
}

@article{georges1996dynamical,
  title = {Dynamical mean-field theory of strongly correlated fermion systems and the limit of infinite dimensions},
  author = {Georges, Antoine and Kotliar, Gabriel and Krauth, Werner and Rozenberg, Marcelo J.},
  journal = {Rev. Mod. Phys.},
  volume = {68},
  issue = {1},
  pages = {13--125},
  numpages = {0},
  year = {1996},
  month = {Jan},
  publisher = {American Physical Society},
  doi = {10.1103/RevModPhys.68.13},
  url = {https://link.aps.org/doi/10.1103/RevModPhys.68.13}
}

@article{tasaki1998nagaoka,
    author = {Tasaki, Hal},
    title = {From Nagaoka's Ferromagnetism to Flat-Band Ferromagnetism and Beyond: An Introduction to Ferromagnetism in the Hubbard Model},
    journal = {Progress of Theoretical Physics},
    volume = {99},
    number = {4},
    pages = {489-548},
    year = {1998},
    month = {04},
    issn = {0033-068X},
    doi = {10.1143/PTP.99.489},
    url = {https://doi.org/10.1143/PTP.99.489},
    eprint = {https://academic.oup.com/ptp/article-pdf/99/4/489/5209512/99-4-489.pdf},
}

@article{imada1998metal,
  title = {Metal-insulator transitions},
  author = {Imada, Masatoshi and Fujimori, Atsushi and Tokura, Yoshinori},
  journal = {Rev. Mod. Phys.},
  volume = {70},
  issue = {4},
  pages = {1039--1263},
  numpages = {0},
  year = {1998},
  month = {Oct},
  publisher = {American Physical Society},
  doi = {10.1103/RevModPhys.70.1039},
  url = {https://link.aps.org/doi/10.1103/RevModPhys.70.1039}
}

@article{assaad1999quantum,
  title={Quantum Monte Carlo simulations of the half-filled two-dimensional Kondo lattice model},
  author={Assaad, FF},
  journal={Physical Review Letters},
  volume={83},
  number={4},
  pages={796},
  year={1999},
  publisher={APS},
  url={https://doi.org/10.1103/PhysRevLett.83.796},
  doi={10.1103/PhysRevLett.83.796}
}

@article{grober2000anomalous,
  title = {Anomalous low-doping phase of the Hubbard model},
  author = {Gr\"ober, C. and Eder, R. and Hanke, W.},
  journal = {Phys. Rev. B},
  volume = {62},
  issue = {7},
  pages = {4336--4352},
  numpages = {0},
  year = {2000},
  month = {Aug},
  publisher = {American Physical Society},
  doi = {10.1103/PhysRevB.62.4336},
  url = {https://link.aps.org/doi/10.1103/PhysRevB.62.4336}
}

@article{senechal2000spectral,
  title={Spectral weight of the Hubbard model through cluster perturbation theory},
  author={S{\'e}n{\'e}chal, D and Perez, Danny and Pioro-Ladriere, M},
  journal={Physical review letters},
  volume={84},
  number={3},
  pages={522},
  year={2000},
  publisher={APS}
}

@article{capponi2001spin,
  title={Spin and charge dynamics of the ferromagnetic and antiferromagnetic two-dimensional half-filled Kondo lattice model},
  author={Capponi, Sylvain and Assaad, FF},
  journal={Physical Review B},
  volume={63},
  number={15},
  pages={155114},
  year={2001},
  publisher={APS},
  url = {https://doi.org/10.1103/PhysRevB.63.155114},
  doi = {10.1103/PhysRevB.63.155114}
}

@article{senechal2002cluster,
  title = {Cluster perturbation theory for Hubbard models},
  author = {S\'en\'echal, David and Perez, Danny and Plouffe, Dany},
  journal = {Phys. Rev. B},
  volume = {66},
  issue = {7},
  pages = {075129},
  numpages = {11},
  year = {2002},
  month = {Aug},
  publisher = {American Physical Society},
  doi = {10.1103/PhysRevB.66.075129},
  url = {https://link.aps.org/doi/10.1103/PhysRevB.66.075129}
}

@article{mancini2004hubbard,
  title={The Hubbard model within the equations of motion approach},
  author={Mancini, Ferdinando and Avella, Adolfo},
  journal={Advances in Physics},
  volume={53},
  number={5-6},
  pages={537--768},
  year={2004},
  publisher={Taylor \& Francis}
}

@article{maier2005quantum,
  title = {Quantum cluster theories},
  author = {Maier, Thomas and Jarrell, Mark and Pruschke, Thomas and Hettler, Matthias H.},
  journal = {Rev. Mod. Phys.},
  volume = {77},
  issue = {3},
  pages = {1027--1080},
  numpages = {0},
  year = {2005},
  month = {Oct},
  publisher = {American Physical Society},
  doi = {10.1103/RevModPhys.77.1027},
  url = {https://link.aps.org/doi/10.1103/RevModPhys.77.1027}
}

@book{ovchinnikov2004hubbard,
  title={Hubbard operators in the theory of strongly correlated electrons},
  author={Ovchinnikov, Sergei G and Val'kov, Valery V},
  year={2004},
  publisher={World Scientific}
}

@article{Doped_SymmetricKondo,
  title = {Symmetric Kondo Lattice States in Doped Strained Twisted Bilayer Graphene},
  author = {Hu, Haoyu and Rai, Gautam and Crippa, Lorenzo and Herzog-Arbeitman, Jonah and C\ifmmode \u{a}\else \u{a}\fi{}lug\ifmmode \u{a}\else \u{a}\fi{}ru, Dumitru and Wehling, Tim and Sangiovanni, Giorgio and Valent\'{\i}, Roser and Tsvelik, Alexei M. and Bernevig, B. Andrei},
  journal = {Phys. Rev. Lett.},
  volume = {131},
  issue = {16},
  pages = {166501},
  numpages = {8},
  year = {2023},
  month = {Oct},
  publisher = {American Physical Society},
  doi = {10.1103/PhysRevLett.131.166501},
  url = {https://link.aps.org/doi/10.1103/PhysRevLett.131.166501}
}

@article{bistritzer2011moire,
author = {Rafi Bistritzer  and Allan H. MacDonald },
title = {Moiré bands in twisted double-layer graphene},
journal = {Proceedings of the National Academy of Sciences},
volume = {108},
number = {30},
pages = {12233-12237},
year = {2011},
doi = {10.1073/pnas.1108174108},
URL = {https://www.pnas.org/doi/abs/10.1073/pnas.1108174108},
eprint = {https://www.pnas.org/doi/pdf/10.1073/pnas.1108174108}}

@article{silant2015dimer,
author={Silant'ev, A. V.},
title={A Dimer in the Extended Hubbard Model},
journal={Russian Physics Journal},
year={2015},
month={Mar},
day={01},
volume={57},
number={11},
pages={1491-1502},
issn={1573-9228},
doi={10.1007/s11182-015-0406-z},
url={https://doi.org/10.1007/s11182-015-0406-z}
}

@article{bercx2017alf,
	title={{The ALF (Algorithms for Lattice Fermions) project release 1.0. Documentation for the auxiliary field quantum Monte Carlo code}},
	author={Martin Bercx and Florian Goth and Johannes S. Hofmann and Fakher F. Assaad},
	journal={SciPost Phys.},
	volume={3},
	pages={013},
	year={2017},
	publisher={SciPost},
	doi={10.21468/SciPostPhys.3.2.013},
	url={https://scipost.org/10.21468/SciPostPhys.3.2.013},
}

@article{cao2018sc,
author={Cao, Yuan
and Fatemi, Valla
and Fang, Shiang
and Watanabe, Kenji
and Taniguchi, Takashi
and Kaxiras, Efthimios
and Jarillo-Herrero, Pablo},
title={Unconventional superconductivity in magic-angle graphene superlattices},
journal={Nature},
year={2018},
month={Apr},
day={01},
volume={556},
number={7699},
pages={43-50},
issn={1476-4687},
doi={10.1038/nature26160},
url={https://doi.org/10.1038/nature26160}
}

@article{catalano2018hubbard,
  title={The Hubbard dimer within the Green's function equation of motion approach},
  author={Catalano, Francesco and Nilsson, Johan},
  journal={arXiv preprint arXiv:1807.07717},
  year={2018}
}

@article{kang2018symmetry,
  title = {Symmetry, Maximally Localized Wannier States, and a Low-Energy Model for Twisted Bilayer Graphene Narrow Bands},
  author = {Kang, Jian and Vafek, Oskar},
  journal = {Phys. Rev. X},
  volume = {8},
  issue = {3},
  pages = {031088},
  numpages = {9},
  year = {2018},
  month = {Sep},
  publisher = {American Physical Society},
  doi = {10.1103/PhysRevX.8.031088},
  url = {https://link.aps.org/doi/10.1103/PhysRevX.8.031088}
}

@article{cao2018correlated,
author={Cao, Yuan
and Fatemi, Valla
and Demir, Ahmet
and Fang, Shiang
and Tomarken, Spencer L.
and Luo, Jason Y.
and Sanchez-Yamagishi, Javier D.
and Watanabe, Kenji
and Taniguchi, Takashi
and Kaxiras, Efthimios
and Ashoori, Ray C.
and Jarillo-Herrero, Pablo},
title={Correlated insulator behaviour at half-filling in magic-angle graphene superlattices},
journal={Nature},
year={2018},
month={Apr},
day={01},
volume={556},
number={7699},
pages={80-84},
issn={1476-4687},
doi={10.1038/nature26154},
url={https://doi.org/10.1038/nature26154}
}

@article{choi2019stm,
author={Choi, Youngjoon
and Kemmer, Jeannette
and Peng, Yang
and Thomson, Alex
and Arora, Harpreet
and Polski, Robert
and Zhang, Yiran
and Ren, Hechen
and Alicea, Jason
and Refael, Gil
and von Oppen, Felix
and Watanabe, Kenji
and Taniguchi, Takashi
and Nadj-Perge, Stevan},
title={Electronic correlations in twisted bilayer graphene near the magic angle},
journal={Nature Physics},
year={2019},
month={Nov},
day={01},
volume={15},
number={11},
pages={1174-1180},
issn={1745-2481},
doi={10.1038/s41567-019-0606-5},
url={https://doi.org/10.1038/s41567-019-0606-5}
}

@article{kerelsky2019stm,
author={Kerelsky, Alexander
and McGilly, Leo J.
and Kennes, Dante M.
and Xian, Lede
and Yankowitz, Matthew
and Chen, Shaowen
and Watanabe, K.
and Taniguchi, T.
and Hone, James
and Dean, Cory
and Rubio, Angel
and Pasupathy, Abhay N.},
title={Maximized electron interactions at the magic angle in twisted bilayer graphene},
journal={Nature},
year={2019},
month={Aug},
day={01},
volume={572},
number={7767},
pages={95-100},
issn={1476-4687},
doi={10.1038/s41586-019-1431-9},
url={https://doi.org/10.1038/s41586-019-1431-9}
}

@article{lu2019superconductors,
author={Lu, Xiaobo
and Stepanov, Petr
and Yang, Wei
and Xie, Ming
and Aamir, Mohammed Ali
and Das, Ipsita
and Urgell, Carles
and Watanabe, Kenji
and Taniguchi, Takashi
and Zhang, Guangyu
and Bachtold, Adrian
and MacDonald, Allan H.
and Efetov, Dmitri K.},
title={Superconductors, orbital magnets and correlated states in magic-angle bilayer graphene},
journal={Nature},
year={2019},
month={Oct},
day={01},
volume={574},
number={7780},
pages={653-657},
issn={1476-4687},
doi={10.1038/s41586-019-1695-0}
}

@article{xie2019stm,
author={Xie, Yonglong
and Lian, Biao
and J{\"a}ck, Berthold
and Liu, Xiaomeng
and Chiu, Cheng-Li
and Watanabe, Kenji
and Taniguchi, Takashi
and Bernevig, B. Andrei
and Yazdani, Ali},
title={Spectroscopic signatures of many-body correlations in magic-angle twisted bilayer graphene},
journal={Nature},
year={2019},
month={Aug},
day={01},
volume={572},
number={7767},
pages={101-105},
issn={1476-4687},
doi={10.1038/s41586-019-1422-x},
url={https://doi.org/10.1038/s41586-019-1422-x}
}

@article{sharpe2019ferro,
author = {Aaron L. Sharpe  and Eli J. Fox  and Arthur W. Barnard  and Joe Finney  and Kenji Watanabe  and Takashi Taniguchi  and M. A. Kastner  and David Goldhaber-Gordon },
title = {Emergent ferromagnetism near three-quarters filling in twisted bilayer graphene},
journal = {Science},
volume = {365},
number = {6453},
pages = {605-608},
year = {2019},
doi = {10.1126/science.aaw3780},
URL = {https://www.science.org/doi/abs/10.1126/science.aaw3780},
eprint = {https://www.science.org/doi/pdf/10.1126/science.aaw3780}}

@article{yankowitz2019tuning,
author = {Matthew Yankowitz  and Shaowen Chen  and Hryhoriy Polshyn  and Yuxuan Zhang  and K. Watanabe  and T. Taniguchi  and David Graf  and Andrea F. Young  and Cory R. Dean },
title = {Tuning superconductivity in twisted bilayer graphene},
journal = {Science},
volume = {363},
number = {6431},
pages = {1059-1064},
year = {2019},
doi = {10.1126/science.aav1910}}

@article{saito2020independent,
author={Saito, Yu
and Ge, Jingyuan
and Watanabe, Kenji
and Taniguchi, Takashi
and Young, Andrea F.},
title={Independent superconductors and correlated insulators in twisted bilayer graphene},
journal={Nature Physics},
year={2020},
month={Sep},
day={01},
volume={16},
number={9},
pages={926-930},
issn={1745-2481},
doi={10.1038/s41567-020-0928-3}
}

@article{stepanov2020untying,
author={Stepanov, Petr
and Das, Ipsita
and Lu, Xiaobo
and Fahimniya, Ali
and Watanabe, Kenji
and Taniguchi, Takashi
and Koppens, Frank H. L.
and Lischner, Johannes
and Levitov, Leonid
and Efetov, Dmitri K.},
title={Untying the insulating and superconducting orders in magic-angle graphene},
journal={Nature},
year={2020},
month={Jul},
day={01},
volume={583},
number={7816},
pages={375-378},
issn={1476-4687},
doi={10.1038/s41586-020-2459-6}
}

@article{xie2020nature,
  title = {Nature of the Correlated Insulator States in Twisted Bilayer Graphene},
  author = {Xie, Ming and MacDonald, A. H.},
  journal = {Phys. Rev. Lett.},
  volume = {124},
  issue = {9},
  pages = {097601},
  numpages = {6},
  year = {2020},
  month = {Mar},
  publisher = {American Physical Society},
  doi = {10.1103/PhysRevLett.124.097601},
  url = {https://link.aps.org/doi/10.1103/PhysRevLett.124.097601}
}

@article{bernevig2021twisted,
  title={Twisted bilayer graphene. III. Interacting Hamiltonian and exact symmetries},
  author={Bernevig, B Andrei and Song, Zhi-Da and Regnault, Nicolas and Lian, Biao},
  journal={Physical Review B},
  volume={103},
  number={20},
  pages={205413},
  year={2021},
  publisher={APS}
}

@article{rozen2021entropic,
  title={Entropic evidence for a Pomeranchuk effect in magic-angle graphene},
  author={Rozen, Asaf and Park, Jeong Min and Zondiner, Uri and Cao, Yuan and Rodan-Legrain, Daniel and Taniguchi, Takashi and Watanabe, Kenji and Oreg, Yuval and Stern, Ady and Berg, Erez and others},
  journal={Nature},
  volume={592},
  number={7853},
  pages={214--219},
  year={2021},
  publisher={Nature Publishing Group UK London},
  doi={10.1038/s41586-021-03319-3},
  url={https://doi.org/10.1038/s41586-021-03319-3}
}

@article{jaoui2022strangemetal,
author={Jaoui, Alexandre
and Das, Ipsita
and Di Battista, Giorgio
and D{\'i}ez-M{\'e}rida, Jaime
and Lu, Xiaobo
and Watanabe, Kenji
and Taniguchi, Takashi
and Ishizuka, Hiroaki
and Levitov, Leonid
and Efetov, Dmitri K.},
title={Quantum critical behaviour in magic-angle twisted bilayer graphene},
journal={Nature Physics},
year={2022},
month={Jun},
day={01},
volume={18},
number={6},
pages={633-638},
issn={1745-2481},
doi={10.1038/s41567-022-01556-5},
url={https://doi.org/10.1038/s41567-022-01556-5}
}

@article{wong2020cascade,
author={Wong, Dillon
and Nuckolls, Kevin P.
and Oh, Myungchul
and Lian, Biao
and Xie, Yonglong
and Jeon, Sangjun
and Watanabe, Kenji
and Taniguchi, Takashi
and Bernevig, B. Andrei
and Yazdani, Ali},
title={Cascade of electronic transitions in magic-angle twisted bilayer graphene},
journal={Nature},
year={2020},
month={Jun},
day={01},
volume={582},
number={7811},
pages={198-202},
issn={1476-4687},
doi={10.1038/s41586-020-2339-0},
url={https://doi.org/10.1038/s41586-020-2339-0}
}

@article{zondiner2020cascade,
author={Zondiner, U.
and Rozen, A.
and Rodan-Legrain, D.
and Cao, Y.
and Queiroz, R.
and Taniguchi, T.
and Watanabe, K.
and Oreg, Y.
and von Oppen, F.
and Stern, Ady
and Berg, E.
and Jarillo-Herrero, P.
and Ilani, S.},
title={Cascade of phase transitions and Dirac revivals in magic-angle graphene},
journal={Nature},
year={2020},
month={Jun},
day={01},
volume={582},
number={7811},
pages={203-208},
issn={1476-4687},
doi={10.1038/s41586-020-2373-y},
url={https://doi.org/10.1038/s41586-020-2373-y}
}

@article{das2021qoscs,
author={Das, Ipsita
and Lu, Xiaobo
and Herzog-Arbeitman, Jonah
and Song, Zhi-Da
and Watanabe, Kenji
and Taniguchi, Takashi
and Bernevig, B. Andrei
and Efetov, Dmitri K.},
title={Symmetry-broken Chern insulators and Rashba-like Landau-level crossings in magic-angle bilayer graphene},
journal={Nature Physics},
year={2021},
month={Jun},
day={01},
volume={17},
number={6},
pages={710-714},
issn={1745-2481},
doi={10.1038/s41567-021-01186-3},
url={https://doi.org/10.1038/s41567-021-01186-3}
}

@article{khalaf2019magic,
  title = {Magic angle hierarchy in twisted graphene multilayers},
  author = {Khalaf, Eslam and Kruchkov, Alex J. and Tarnopolsky, Grigory and Vishwanath, Ashvin},
  journal = {Phys. Rev. B},
  volume = {100},
  issue = {8},
  pages = {085109},
  numpages = {9},
  year = {2019},
  month = {Aug},
  publisher = {American Physical Society},
  doi = {10.1103/PhysRevB.100.085109},
  url = {https://link.aps.org/doi/10.1103/PhysRevB.100.085109}
}

@article{ahn2019failure,
  title = {Failure of Nielsen-Ninomiya Theorem and Fragile Topology in Two-Dimensional Systems with Space-Time Inversion Symmetry: Application to Twisted Bilayer Graphene at Magic Angle},
  author = {Ahn, Junyeong and Park, Sungjoon and Yang, Bohm-Jung},
  journal = {Phys. Rev. X},
  volume = {9},
  issue = {2},
  pages = {021013},
  numpages = {26},
  year = {2019},
  month = {Apr},
  publisher = {American Physical Society},
  doi = {10.1103/PhysRevX.9.021013},
  url = {https://link.aps.org/doi/10.1103/PhysRevX.9.021013}
}

@article{song2019all,
  title = {All Magic Angles in Twisted Bilayer Graphene are Topological},
  author = {Song, Zhida and Wang, Zhijun and Shi, Wujun and Li, Gang and Fang, Chen and Bernevig, B. Andrei},
  journal = {Phys. Rev. Lett.},
  volume = {123},
  issue = {3},
  pages = {036401},
  numpages = {6},
  year = {2019},
  month = {Jul},
  publisher = {American Physical Society},
  doi = {10.1103/PhysRevLett.123.036401},
  url = {https://link.aps.org/doi/10.1103/PhysRevLett.123.036401}
}

@article{po2019faithful,
  title = {Faithful tight-binding models and fragile topology of magic-angle bilayer graphene},
  author = {Po, Hoi Chun and Zou, Liujun and Senthil, T. and Vishwanath, Ashvin},
  journal = {Phys. Rev. B},
  volume = {99},
  issue = {19},
  pages = {195455},
  numpages = {16},
  year = {2019},
  month = {May},
  publisher = {American Physical Society},
  doi = {10.1103/PhysRevB.99.195455},
  url = {https://link.aps.org/doi/10.1103/PhysRevB.99.195455}
}

@article{tarnopolsky2019origin,
  title = {Origin of Magic Angles in Twisted Bilayer Graphene},
  author = {Tarnopolsky, Grigory and Kruchkov, Alex Jura and Vishwanath, Ashvin},
  journal = {Phys. Rev. Lett.},
  volume = {122},
  issue = {10},
  pages = {106405},
  numpages = {6},
  year = {2019},
  month = {Mar},
  publisher = {American Physical Society},
  doi = {10.1103/PhysRevLett.122.106405},
  url = {https://link.aps.org/doi/10.1103/PhysRevLett.122.106405}
}

@article{bultinck2020ground,
  title = {Ground State and Hidden Symmetry of Magic-Angle Graphene at Even Integer Filling},
  author = {Bultinck, Nick and Khalaf, Eslam and Liu, Shang and Chatterjee, Shubhayu and Vishwanath, Ashvin and Zaletel, Michael P.},
  journal = {Phys. Rev. X},
  volume = {10},
  issue = {3},
  pages = {031034},
  numpages = {13},
  year = {2020},
  month = {Aug},
  publisher = {American Physical Society},
  doi = {10.1103/PhysRevX.10.031034},
  url = {https://link.aps.org/doi/10.1103/PhysRevX.10.031034}
}

@article{kwan2021kekule,
  title = {Kekul\'e Spiral Order at All Nonzero Integer Fillings in Twisted Bilayer Graphene},
  author = {Kwan, Y. H. and Wagner, G. and Soejima, T. and Zaletel, M. P. and Simon, S. H. and Parameswaran, S. A. and Bultinck, N.},
  journal = {Phys. Rev. X},
  volume = {11},
  issue = {4},
  pages = {041063},
  numpages = {23},
  year = {2021},
  month = {Dec},
  publisher = {American Physical Society},
  doi = {10.1103/PhysRevX.11.041063},
  url = {https://link.aps.org/doi/10.1103/PhysRevX.11.041063}
}

@article{khalaf2021charged,
author = {Eslam Khalaf  and Shubhayu Chatterjee  and Nick Bultinck  and Michael P. Zaletel  and Ashvin Vishwanath },
title = {Charged skyrmions and topological origin of superconductivity in magic-angle graphene},
journal = {Science Advances},
volume = {7},
number = {19},
pages = {eabf5299},
year = {2021},
doi = {10.1126/sciadv.abf5299},
URL = {https://www.science.org/doi/abs/10.1126/sciadv.abf5299}}

@article{saito2021isospin,
  title={Isospin Pomeranchuk effect in twisted bilayer graphene},
  author={Saito, Yu and Yang, Fangyuan and Ge, Jingyuan and Liu, Xiaoxue and Taniguchi, Takashi and Watanabe, Kenji and Li, JIA and Berg, Erez and Young, Andrea F},
  journal={Nature},
  volume={592},
  number={7853},
  pages={220--224},
  year={2021},
  publisher={Nature Publishing Group UK London},
  doi = {10.1038/s41586-021-03409-2},
  url = {https://doi.org/10.1038/s41586-021-03409-2}
}

@article{shavit2021theory,
  title = {Theory of Correlated Insulators and Superconductivity in Twisted Bilayer Graphene},
  author = {Shavit, Gal and Berg, Erez and Stern, Ady and Oreg, Yuval},
  journal = {Phys. Rev. Lett.},
  volume = {127},
  issue = {24},
  pages = {247703},
  numpages = {6},
  year = {2021},
  month = {Dec},
  publisher = {American Physical Society},
  doi = {10.1103/PhysRevLett.127.247703},
  url = {https://link.aps.org/doi/10.1103/PhysRevLett.127.247703}
}

@article{wu2021chern,
author={Wu, Shuang
and Zhang, Zhenyuan
and Watanabe, K.
and Taniguchi, T.
and Andrei, Eva Y.},
title={Chern insulators, van Hove singularities and topological flat bands in magic-angle twisted bilayer graphene},
journal={Nature Materials},
year={2021},
month={Apr},
day={01},
volume={20},
number={4},
pages={488-494},
issn={1476-4660},
doi={10.1038/s41563-020-00911-2},
url={https://doi.org/10.1038/s41563-020-00911-2}
}

@article{arovas2022hubbard,
   author = "Arovas, Daniel P. and Berg, Erez and Kivelson, Steven A. and Raghu, Srinivas",
   title = "The Hubbard Model", 
   journal= "Annual Review of Condensed Matter Physics",
   year = "2022",
   volume = "13",
   number = "Volume 13, 2022",
   pages = "239-274",
   doi = "https://doi.org/10.1146/annurev-conmatphys-031620-102024",
   publisher = "Annual Reviews"
  }

@article{wagner2022global,
  title = {Global Phase Diagram of the Normal State of Twisted Bilayer Graphene},
  author = {Wagner, Glenn and Kwan, Yves H. and Bultinck, Nick and Simon, Steven H. and Parameswaran, S. A.},
  journal = {Phys. Rev. Lett.},
  volume = {128},
  issue = {15},
  pages = {156401},
  numpages = {7},
  year = {2022},
  month = {Apr},
  publisher = {American Physical Society},
  doi = {10.1103/PhysRevLett.128.156401},
  url = {https://link.aps.org/doi/10.1103/PhysRevLett.128.156401}
}

@article{qin2022hubbard,
   author = "Qin, Mingpu and Schäfer, Thomas and Andergassen, Sabine and Corboz, Philippe and Gull, Emanuel",
   title = "The Hubbard Model: A Computational Perspective", 
   journal= "Annual Review of Condensed Matter Physics",
   year = "2022",
   volume = "13",
   number = "Volume 13, 2022",
   pages = "275-302",
   doi = "https://doi.org/10.1146/annurev-conmatphys-090921-033948",
   url = "https://www.annualreviews.org/content/journals/10.1146/annurev-conmatphys-090921-033948",
   publisher = "Annual Reviews",
   issn = "1947-5462"}

@article{song2022magic,
  title = {Magic-Angle Twisted Bilayer Graphene as a Topological Heavy Fermion Problem},
  author = {Song, Zhi-Da and Bernevig, B. Andrei},
  journal = {Phys. Rev. Lett.},
  volume = {129},
  issue = {4},
  pages = {047601},
  numpages = {10},
  year = {2022},
  month = {Jul},
  publisher = {American Physical Society},
  doi = {10.1103/PhysRevLett.129.047601},
  url = {https://link.aps.org/doi/10.1103/PhysRevLett.129.047601}
}

@article{kohno2022emergence,
  title = {Emergence of electronic modes by doping Kondo insulators in the Kondo lattice and periodic Anderson models},
  author = {Kohno, Masanori},
  journal = {Phys. Rev. B},
  volume = {105},
  issue = {15},
  pages = {155134},
  numpages = {14},
  year = {2022},
  month = {Apr},
  publisher = {American Physical Society},
  doi = {10.1103/PhysRevB.105.155134},
  url = {https://link.aps.org/doi/10.1103/PhysRevB.105.155134}
}

@article{datta2023heavy,
author={Datta, Anushree
and Calder{\'o}n, M. J.
and Camjayi, A.
and Bascones, E.},
title={Heavy quasiparticles and cascades without symmetry breaking in twisted bilayer graphene},
journal={Nature Communications},
year={2023},
month={Aug},
day={18},
volume={14},
number={1},
pages={5036},
issn={2041-1723},
doi={10.1038/s41467-023-40754-4},
url={https://doi.org/10.1038/s41467-023-40754-4}
}

@book{Correl21,
 title={Simulating correlations with computers. Lecture notes of the autumn school on correlated electrons 2021},
 publisher={Institute for Advanced Simulation},
 volume={11},
 ISSN={2192-8525},
 DOI={10.18154/RWTH-2021-11799},
 author={Pavarini, Eva and Koch, Erik and Forschungszentrum Jülich GmbH (Germany). Institute for Advanced Simulation},
 year={2021} }

@article{tian2023evidence,
author={Tian, Haidong
and Gao, Xueshi
and Zhang, Yuxin
and Che, Shi
and Xu, Tianyi
and Cheung, Patrick
and Watanabe, Kenji
and Taniguchi, Takashi
and Randeria, Mohit
and Zhang, Fan
and Lau, Chun Ning
and Bockrath, Marc W.},
title={Evidence for Dirac flat band superconductivity enabled by quantum geometry},
journal={Nature},
year={2023},
month={Feb},
day={01},
volume={614},
number={7948},
pages={440-444},
issn={1476-4687},
doi={10.1038/s41586-022-05576-2},
url={https://doi.org/10.1038/s41586-022-05576-2}
}

@article{kwan2024electron,
  title = {Electron-phonon coupling and competing Kekul\'e orders in twisted bilayer graphene},
  author = {Kwan, Yves H. and Wagner, Glenn and Bultinck, Nick and Simon, Steven H. and Berg, Erez and Parameswaran, S. A.},
  journal = {Phys. Rev. B},
  volume = {110},
  issue = {8},
  pages = {085160},
  numpages = {9},
  year = {2024},
  month = {Aug},
  publisher = {American Physical Society},
  doi = {10.1103/PhysRevB.110.085160},
  url = {https://link.aps.org/doi/10.1103/PhysRevB.110.085160}
}

@article{haurie2024bands,
doi = {10.1088/1361-648X/ad1e07},
url = {https://doi.org/10.1088/1361-648X/ad1e07},
year = {2024},
month = {mar},
publisher = {IOP Publishing},
volume = {36},
number = {25},
pages = {255601},
author = {Haurie, L and Grandadam, M and Pangburn, E and Banerjee, A and Burdin, S and Pépin, C},
title = {Bands renormalization and superconductivity in the strongly correlated Hubbard model using composite operators method},
journal = {Journal of Physics: Condensed Matter}
}

@article{rai2024dynamical,
  title = {Dynamical Correlations and Order in Magic-Angle Twisted Bilayer Graphene},
  author = {Rai, Gautam and Crippa, Lorenzo and C\ifmmode \u{a}\else \u{a}\fi{}lug\ifmmode \u{a}\else \u{a}\fi{}ru, Dumitru and Hu, Haoyu and Paoletti, Francesca and de' Medici, Luca and Georges, Antoine and Bernevig, B. Andrei and Valent\'{\i}, Roser and Sangiovanni, Giorgio and Wehling, Tim},
  journal = {Phys. Rev. X},
  volume = {14},
  issue = {3},
  pages = {031045},
  numpages = {22},
  year = {2024},
  month = {Sep},
  publisher = {American Physical Society},
  doi = {10.1103/PhysRevX.14.031045},
  url = {https://link.aps.org/doi/10.1103/PhysRevX.14.031045}
}

@article{banerjee2025probing,
  title = {Probing quasiparticle excitations in a doped Mott insulator via Friedel oscillations},
  author = {Banerjee, Anurag and Pangburn, Emile and P\'epin, Catherine and Bena, Cristina},
  journal = {Phys. Rev. B},
  volume = {112},
  issue = {16},
  pages = {165112},
  numpages = {12},
  year = {2025},
  month = {Oct},
  publisher = {American Physical Society},
  doi = {10.1103/fbq7-792q},
  url = {https://link.aps.org/doi/10.1103/fbq7-792q}
}

@article{banerjee2025charge,
  title = {Charge density wave solutions of the Hubbard model in the composite operator formalism},
  author = {Banerjee, Anurag and Pangburn, Emile and Mahato, Chiranjit and Ghosal, Amit and P\'epin, Catherine},
  journal = {Phys. Rev. B},
  volume = {111},
  issue = {16},
  pages = {165123},
  numpages = {15},
  year = {2025},
  month = {Apr},
  publisher = {American Physical Society},
  doi = {10.1103/PhysRevB.111.165123},
  url = {https://link.aps.org/doi/10.1103/PhysRevB.111.165123}
}

@article{crippa2025dynamical,
  title={Dynamical correlation effects in twisted bilayer graphene under strain and lattice relaxation},
  author={Crippa, Lorenzo and Rai, Gautam and C{\u{a}}lug{\u{a}}ru, Dumitru and Hu, Haoyu and Herzog-Arbeitman, Jonah and Bernevig, B Andrei and Valent{\'\i}, Roser and Sangiovanni, Giorgio and Wehling, Tim},
  journal={arXiv preprint arXiv:2509.19436},
  year={2025}
}

@article{cualuguaru2025obtaining,
  title={Obtaining the Spectral Function of Moir$\backslash$'e Graphene Heavy-Fermions Using Iterative Perturbation Theory},
  author={C{\u{a}}lug{\u{a}}ru, Dumitru and Hu, Haoyu and Crippa, Lorenzo and Rai, Gautam and Regnault, Nicolas and Wehling, Tim O and Valent{\'\i}, Roser and Sangiovanni, Giorgio and Bernevig, B Andrei},
  journal={arXiv preprint arXiv:2509.18256},
  year={2025}
}

@article{lau2025topological,
  title={Topological mixed valence model for twisted bilayer graphene},
  author={Lau, Liam LH and Coleman, Piers},
  journal={Physical Review X},
  volume={15},
  number={2},
  pages={021028},
  year={2025},
  publisher={APS},
  doi = {10.1103/PhysRevX.15.021028},
  url = {https://doi.org/10.1103/PhysRevX.15.021028} 
}

@article{JarrellBaye,
title = {Bayesian inference and the analytic continuation of imaginary-time quantum Monte Carlo data},
journal = {Physics Reports},
volume = {269},
number = {3},
pages = {133-195},
year = {1996},
issn = {0370-1573},
doi = {https://doi.org/10.1016/0370-1573(95)00074-7},
url = {https://www.sciencedirect.com/science/article/pii/0370157395000747},
author = {Mark Jarrell and J.E. Gubernatis}
}

@article{waschitz2026momentum,
  title = {Momentum-Resolved Spectroscopy of Superconductivity with the Quantum Twisting Microscope},
  author = {Waschitz, Yuval and Stern, Ady and Oreg, Yuval},
  journal = {Phys. Rev. Lett.},
  volume = {136},
  issue = {15},
  pages = {156501},
  numpages = {8},
  year = {2026},
  month = {Apr},
  publisher = {American Physical Society},
  doi = {10.1103/62rf-8m43},
  url = {https://link.aps.org/doi/10.1103/62rf-8m43}
}

@article{SandvikPRB,
  title = {Stochastic method for analytic continuation of quantum Monte Carlo data},
  author = {Sandvik, Anders W.},
  journal = {Phys. Rev. B},
  volume = {57},
  issue = {17},
  pages = {10287--10290},
  numpages = {0},
  year = {1998},
  month = {May},
  publisher = {American Physical Society},
  doi = {10.1103/PhysRevB.57.10287},
  url = {https://link.aps.org/doi/10.1103/PhysRevB.57.10287}
}

@article{wei2026tunneling,
  title = {Tunneling spectroscopy of two-dimensional superconductors with the quantum twisting microscope},
  author = {Wei, Nemin and von Oppen, Felix and Glazman, Leonid I.},
  journal = {Phys. Rev. B},
  volume = {113},
  issue = {6},
  pages = {064502},
  numpages = {16},
  year = {2026},
  month = {Feb},
  publisher = {American Physical Society},
  doi = {10.1103/qsx1-7zmy},
  url = {https://link.aps.org/doi/10.1103/qsx1-7zmy}
}

@article{Nevanlinna,
  title = {Nevanlinna Analytical Continuation},
  author = {Fei, Jiani and Yeh, Chia-Nan and Gull, Emanuel},
  journal = {Phys. Rev. Lett.},
  volume = {126},
  issue = {5},
  pages = {056402},
  numpages = {6},
  year = {2021},
  month = {Feb},
  publisher = {American Physical Society},
  doi = {10.1103/PhysRevLett.126.056402},
  url = {https://link.aps.org/doi/10.1103/PhysRevLett.126.056402}
}

@Article{vonderLinden,
author={von der Linden, W.},
title={Maximum-entropy data analysis},
journal={Applied Physics A},
year={1995},
month={Feb},
day={01},
volume={60},
number={2},
pages={155-165},
issn={1432-0630},
doi={10.1007/BF01538241},
url={https://doi.org/10.1007/BF01538241}
}

@article{SHAO20231,
title = {Progress on stochastic analytic continuation of quantum Monte Carlo data},
journal = {Physics Reports},
volume = {1003},
pages = {1-88},
year = {2023},
note = {Progress on stochastic analytic continuation of quantum Monte Carlo data},
issn = {0370-1573},
doi = {https://doi.org/10.1016/j.physrep.2022.11.002},
url = {https://www.sciencedirect.com/science/article/pii/S0370157322003921},
author = {Hui Shao and Anders W. Sandvik}
}

@article{ledwith2025nonlocal,
  title = {Nonlocal Moments and Mott Semimetal in the Chern Bands of Twisted Bilayer Graphene},
  author = {Ledwith, Patrick J. and Dong, Junkai and Vishwanath, Ashvin and Khalaf, Eslam},
  journal = {Phys. Rev. X},
  volume = {15},
  issue = {2},
  pages = {021087},
  numpages = {40},
  year = {2025},
  month = {Jun},
  publisher = {American Physical Society},
  doi = {10.1103/PhysRevX.15.021087},
  url = {https://link.aps.org/doi/10.1103/PhysRevX.15.021087}
}

@article{ledwith2025exotic,
  title={Exotic carriers from concentrated topology: Dirac trions as the origin of the missing spectral weight in twisted bilayer graphene},
  author={Ledwith, Patrick J and Vishwanath, Ashvin and Khalaf, Eslam},
  journal={arXiv preprint arXiv:2505.08779},
  year={2025}
}

@article{hu2025projected,
  title={Projected and Solvable Topological Heavy Fermion Model of Twisted Bilayer Graphene},
  author={Hu, Haoyu and Song, Zhi-Da and Bernevig, B Andrei},
  journal={arXiv preprint arXiv:2502.14039},
  year={2025}
}

@article{mironov2025dimer,
  title={Dimer in the Hubbard model. Exact and approximate solutions},
  author={Mironov, Gennadiy Ivanovich},
  journal={arXiv preprint arXiv:2504.19680},
  year={2025}
}

@article{merino2025interplay,
  title={Interplay between light and heavy electron bands in magic-angle twisted bilayer graphene},
  author={Merino, Rafael Luque and C{\u{a}}lug{\u{a}}ru, Dumitru and Hu, Haoyu and D{\'\i}ez-M{\'e}rida, Jaime and D{\'\i}ez-Carl{\'o}n, Andr{\'e}s and Taniguchi, Takashi and Watanabe, Kenji and Seifert, Paul and Bernevig, B Andrei and Efetov, Dmitri K},
  journal={Nature Physics},
  volume={21},
  number={7},
  pages={1078--1084},
  year={2025},
  publisher={Nature Publishing Group UK London},
  doi={10.1038/s41567-025-02912-x},
  url={https://doi.org/10.1038/s41567-025-02912-x}
}

@article{wei2025theory,
  title = {Theory of plasmon spectroscopy with the quantum twisting microscope},
  author = {Wei, Nemin and Guinea, Francisco and von Oppen, Felix and Glazman, Leonid I.},
  journal = {Phys. Rev. B},
  volume = {112},
  issue = {15},
  pages = {155157},
  numpages = {19},
  year = {2025},
  month = {Oct},
  publisher = {American Physical Society},
  doi = {10.1103/v8zx-1vhr},
  url = {https://link.aps.org/doi/10.1103/v8zx-1vhr}
}

@article{xiao2025interacting,
  title={The Interacting Energy Bands of Magic Angle Twisted Bilayer Graphene Revealed by the Quantum Twisting Microscope},
  author={Xiao, J and Inbar, A and Birkbeck, J and Gershon, N and Zamir, Y and Taniguchi, T and Watanabe, K and Berg, E and Ilani, S},
  journal={arXiv preprint arXiv:2506.20738},
  year={2025}
}

@article{Pangburn_Ringstates,
  title = {Impurity-induced Mott ring states and Mott zeros ring states in the Hubbard operator formalism},
  author = {Pangburn, Emile and Banerjee, Anurag and P\'epin, Catherine and Bena, Cristina},
  journal = {Phys. Rev. B},
  volume = {112},
  issue = {12},
  pages = {125157},
  numpages = {20},
  year = {2025},
  month = {Sep},
  publisher = {American Physical Society},
  doi = {10.1103/lysb-n9zp},
  url = {https://link.aps.org/doi/10.1103/lysb-n9zp}
}

@article{wei2025dirac,
  title={Dirac-point spectroscopy of flat-band systems with the quantum twisting microscope},
  author={Wei, Nemin and von Oppen, Felix and Glazman, Leonid I},
  journal={Physical Review B},
  volume={111},
  number={8},
  pages={085128},
  year={2025},
  publisher={APS},
  url={ https://doi.org/10.1103/PhysRevB.111.085128},
  doi = {10.1103/PhysRevB.111.085128}
}

@article{Pangburn_Layer,
  title = {Spontaneous layer selective Mott phase in the bilayer Hubbard model},
  author = {Pangburn, Emile and Haurie, Louis and Burdin, S\'ebastien and P\'epin, Catherine and Banerjee, Anurag},
  journal = {Phys. Rev. B},
  volume = {110},
  issue = {23},
  pages = {235142},
  numpages = {14},
  year = {2024},
  month = {Dec},
  publisher = {American Physical Society},
  doi = {10.1103/PhysRevB.110.235142},
  url = {https://link.aps.org/doi/10.1103/PhysRevB.110.235142}
}

@article{Pangburn_TMZ,
  title = {Topological charge excitations and Green's function zeros in paramagnetic Mott insulators},
  author = {Pangburn, Emile and P\'epin, Catherine and Banerjee, Anurag},
  journal = {Phys. Rev. B},
  volume = {112},
  issue = {8},
  pages = {085105},
  numpages = {18},
  year = {2025},
  month = {Aug},
  publisher = {American Physical Society},
  doi = {10.1103/4k4y-4hj4},
  url = {https://link.aps.org/doi/10.1103/4k4y-4hj4}
}

@article{Mixed_Valence,
  title = {Mixed-valence Mott insulator and composite excitation in twisted bilayer graphene},
  author = {Zhao, Jing-Yu and Zhou, Boran and Zhang, Ya-Hui},
  journal = {Phys. Rev. B},
  volume = {112},
  issue = {23},
  pages = {235144},
  numpages = {12},
  year = {2025},
  month = {Dec},
  publisher = {American Physical Society},
  doi = {10.1103/kts3-81nk},
  url = {https://link.aps.org/doi/10.1103/kts3-81nk}
}

@misc{zhou2026symmetric,
      title={Symmetric topological Mott insulator and Mott semimetal}, 
      author={Boran Zhou and Ya-Hui Zhang},
      year={2026},
      eprint={2601.02485},
      archivePrefix={arXiv},
      primaryClass={cond-mat.str-el},
      url={https://arxiv.org/abs/2601.02485}, 
}

@article{zhao2025topological,
  title = {Topological Mott localization and pseudogap metal in twisted bilayer graphene},
  author = {Zhao, Jing-Yu and Zhou, Boran and Zhang, Ya-Hui},
  journal = {Phys. Rev. B},
  volume = {112},
  issue = {8},
  pages = {085107},
  numpages = {25},
  year = {2025},
  month = {Aug},
  publisher = {American Physical Society},
  doi = {10.1103/9n8v-7rx2},
  url = {https://link.aps.org/doi/10.1103/9n8v-7rx2}
}

@article{zhao2025resonating,
  title={Resonating-valence-bond superconductor from small Fermi surface in twisted bilayer graphene},
  author={Zhao, Jing-Yu and Zhang, Ya-Hui},
  journal={arXiv preprint arXiv:2510.26801},
  year={2025}
}

@article{Po2018,
author = {Po, Hoi Chun and Zou, Liujun and Vishwanath, Ashvin and Senthil, T.},
doi = {10.1103/PhysRevX.8.031089},
issn = {2160-3308},
journal = {Physical Review X},
month = {sep},
number = {3},
pages = {031089},
title = {{Origin of Mott Insulating Behavior and Superconductivity in Twisted Bilayer Graphene}},
url = {https://link.aps.org/doi/10.1103/PhysRevX.8.031089},
volume = {8},
year = {2018}
}

@article{zhang2025heavy,
  title={Heavy fermions, mass renormalization and local moments in magic-angle twisted bilayer graphene via planar tunneling spectroscopy},
  author={Zhang, Zhenyuan and Wu, Shuang and C{\u{a}}lug{\u{a}}ru, Dumitru and Hu, Haoyu and Taniguchi, Takashi and Wanatabe, Kenji and Bernevig, Andrei B and Andrei, Eva Y},
  journal={arXiv preprint arXiv:2503.17875},
  year={2025}
}

@misc{vituri2026controlledloopexpansiontopological,
      title={Controlled Loop Expansion for the Topological Heavy Fermion Model}, 
      author={Yaar Vituri and Erez Berg},
      year={2026},
      eprint={2604.14278},
      archivePrefix={arXiv},
      primaryClass={cond-mat.str-el},
      url={https://arxiv.org/abs/2604.14278}, 
}

@misc{hu2026twistedbilayergraphenelifetimes,
      title={Twisted Bilayer Graphene Lifetimes At Integer Fillings: An Analytic Result}, 
      author={Haoyu Hu and Yuelin Shao and Lorenzo Crippa and Dumitru Călugăru and Giorgio Sangiovanni and Tim Wehling and Leonid I. Glazman and B. Andrei Bernevig},
      year={2026},
      eprint={2604.14303},
      archivePrefix={arXiv},
      primaryClass={cond-mat.str-el},
      url={https://arxiv.org/abs/2604.14303}, 
}

@misc{wei2026lifetimespectralfunctiontopological,
      title={Lifetime and spectral function of topological heavy fermions}, 
      author={Nemin Wei and Felix von Oppen and Leonid I. Glazman},
      year={2026},
      eprint={2604.14369},
      archivePrefix={arXiv},
      primaryClass={cond-mat.str-el},
      url={https://arxiv.org/abs/2604.14369}, 
}

@article{ma2026green,
  title={Green's Function-Free Formalism of Projective Truncation Approximation},
  author={Ma, Kou-Han and Wu, Yue-Hong and Tong, Ning-Hua},
  journal={arXiv preprint arXiv:2605.19404},
  year={2026}
}

@article{nosov2026controlled,
  title={Controlled expansion for correlated electrons with concentrated kinematics},
  author={Nosov, Pavel A and Khalaf, Eslam and Ledwith, Patrick},
  journal={arXiv preprint arXiv:2605.20171},
  year={2026}
}
\end{document}